\documentclass[a4paper,11pt]{article}

\pdfoutput=1

\usepackage{jcappub}
\usepackage[english]{babel}
\usepackage{amsfonts}
\usepackage{txfonts}
\usepackage{graphicx}
\usepackage{amssymb}
\usepackage[T1]{fontenc}
\usepackage{ae,aecompl}
\usepackage{fancyhdr}
\usepackage{multicol}
\usepackage{layout}
\usepackage{graphicx}
\usepackage{multirow}
\usepackage{color}
\usepackage{times}
\usepackage{textcomp}
\usepackage{cprotect}
\usepackage[linesnumbered,boxed,commentsnumbered]{algorithm2e}
\usepackage{algpseudocode}

\graphicspath{{./Plots/}}

\newif\ifAMStwofonts
\AMStwofontstrue

\title{On the implementation of the spherical collapse model for dark energy models}

\author[a,1]{Francesco Pace,%
\note{Both authors contributed equally to this work.}}
\author[b,1]{Sven Meyer}
\author[b]{and Matthias Bartelmann}

\affiliation[a]{Jodrell Bank Centre for Astrophysics, School of Physics and Astronomy, The University of Manchester, 
Manchester, M13 9PL, United Kingdom}
\affiliation[b]{Zentrum f{\"u}r Astronomie der Universit{\"a}t Heidelberg, Institut f{\"u}r theoretische Astrophysik, 
Philosophenweg 12, D-69120, Heidelberg, Germany}

\emailAdd{francesco.pace@manchester.ac.uk}
\emailAdd{sven.meyer@uni-heidelberg.de}
\emailAdd{bartelmann@uni-heidelberg.de}

\abstract{
In this work we review the theory of the spherical collapse model and critically analyse the aspects of the numerical 
implementation of its fundamental equations. By extending a recent work by \cite{Herrera2017}, we show how different 
aspects, such as the initial integration time, the definition of constant infinity and the criterion for the 
extrapolation method (how close the inverse of the overdensity has to be to zero at the collapse time) can lead to an 
erroneous estimation (a few per mill error which translates to a few percent in the mass function) of the key quantity 
in the spherical collapse model: the linear critical overdensity $\delta_{\rm c}$, which plays a crucial role for the 
mass function of halos. 
We provide a better recipe to adopt in designing a code suitable to a generic smooth dark energy model and we compare 
our numerical results with analytic predictions for the EdS and the $\Lambda$CDM models. We further discuss the 
evolution of $\delta_{\rm c}$ for selected classes of dark energy models as a general test of the robustness of our 
implementation. We finally outline which modifications need to be taken into account to extend the code to more general 
classes of models, such as clustering dark energy models and non-minimally coupled models.
}

\keywords{Large scale structure of the universe - dark energy theory - cosmological perturbation theory}

\arxivnumber{1708.02477}

\begin{document}

\label{firstpage}

\maketitle
\flushbottom

\section{Introduction}
Since the discovery of the accelerated expansion of the Universe \citep[e.g.][]{Riess1998,Perlmutter1999}, later 
confirmed by a plethora of other probes \citep[e.g.][]{Cole2005,Komatsu2011,Planck2016_XIII,Planck2016_XIV}, much 
research in cosmology has been devoted to its theoretical explanation. While at the moment a complete physical 
understanding is lacking, a general concordance cosmological model has been designed. According to it, the Universe is 
spatially flat, its expansion is homogeneous in all directions and the main source of the gravitational potential is 
due to the cold dark matter (CDM) component, whose effects are, as for our understanding, only of gravitational 
origin. While CDM adds up to 28\% of the total energy budget of the Universe and standard matter, dubbed generically 
baryons, amounts to about 5\%, the rest (approximately 67\%) is the fluid responsible for the accelerated expansion of 
the Universe \citep{Planck2016_XIII}. 
In its simplest version, according to the concordance cosmological model, this is represented by the cosmological 
constant $\Lambda$, a fluid with equation of state $w_{\Lambda}=P_{\Lambda}/(\rho_{\Lambda}c^2)=-1$ and 
constant throughout the cosmic time. This model is commonly referred to as $\Lambda$CDM model.

The $\Lambda$CDM cosmology is a simple yet very powerful model able to explain the wealth of data available. It 
relies on the symmetries of the Friedmann-Robertson-Walker metric and on General Relativity as the theory describing 
the gravitational interaction. Nevertheless, the $\Lambda$CDM model is affected by severe conceptual and theoretical 
problems \citep{Weinberg1989,Nobbenhuis2006,Polchinski2006,Bousso2008,Martin2012,Burgess2013,Velten2014b}. 
On the large scales probed so far, the model is very successful, but the same cannot be said for small scales
\citep{DelPopolo2017a,Bullock2017}, even if on such scales, gas physics becomes very important.

Lacking a conclusive evidence about the cosmological constant, other models have been investigated and they can be 
divided into two main groups. The first one still relies on General Relativity and includes the so-called dark 
energy (DE) models. These models are characterised by an equation of state which, in principle, can be an arbitrary 
function of time (with the condition $w<-1/3$ to achieve an accelerated expansion).
\footnote{More precisely one should consider the effective equation of state obtained by considering all the species. 
In this case $w_{\rm eff}\lesssim -1/2$.} 
The equation of state is therefore regarded as a free function depending on some parameters which will be fitted to 
data. The simplest and most used model is the CPL parametrization \citep{Chevallier2001,Linder2003} and the equation of 
state $w(a)$ is a linear function of the scale factor. 
These models can be easily studied and put on firm theoretical basis in the framework of the scalar fields. When doing 
this, it is then necessary to specify the form of the self-interacting potential $V(\phi)$. Our ignorance on the 
equation of state is then translated into our ignorance of the potential. We refer to \cite{Battye2016} and references 
therein for a list and a comparison about the background properties of quintessence and $k$-essence models.\\
A second class describes the so-called modified gravity models where the Lagrangian for the gravitational sector is 
different from the General Relativistic one. Notable examples are Brans-Dicke theory 
\citep{Brans1961,DeFelice2010a}, Kinetic Gravity Braiding (KGB) \citep{Deffayet2010,Pujolas2011} and $f(R)$ models 
\citep{Silvestri2009,Sotiriou2010,DeFelice2010,Nojiri2011,Nojiri2017}. For recent reviews, we refer to 
\cite{Clifton2012,Tsujikawa2013a,Tsujikawa2013b,Joyce2015,Joyce2016,Koyama2016,Sami2016,BeltranJimenez2017}. \\
The majority of dark energy and modified gravity models can be seen as particular subclasses of the Horndeski models 
\citep{Horndeski1974,Deffayet2011,Kobayashi2011}, which represent the most general models with second order equations 
of motion.

It is always possible to construct designer dark energy and modified gravity models where the background is, by 
construction, fully described by the equation of state \citep{He2012,He2013,Battye2016}. When this is chosen to 
be $w=-1$, at the background level the models are identical to the $\Lambda$CDM model. To distinguish them, it is 
therefore necessary to study the evolution of perturbations, in particular on non-linear scales, such as clusters 
\citep{Schmidt2009a}. Their abundance is strictly related to the underlying cosmological model 
\citep{Sunyaev1980b,Majumdar2004,Diego2004,Fang2007,Abramo2009a,Angrick2009} and has direct influence 
on weak lensing peak counts \citep{Maturi2010,Maturi2011,Lin2014,Reischke2016}. In particular, it is sensitive to the 
normalization of the matter power spectrum $\sigma_8$ and the matter density parameter $\Omega_{\rm m}$ 
\citep{Angrick2015} and to the evolution of the dark energy equation of state 
\citep{Holder2001,Haiman2001,Weller2002,Majumdar2003}.

A simple, but very powerful tool to study the non-linear evolution of structures is the spherical collapse model. This 
analytic approach was introduced by \cite{Tomita1969,Gunn1972} and lately extended and improved by many authors 
\citep{Fillmore1984,Bertschinger1985,Hoffman1985,Ryden1987,Lahav1991,AvilaReese1998,Subramanian2000,Ascasibar2004,
Mota2004,Williams2004,Shi2016}. 
More recent works took into account extensions to smooth 
\citep{Basilakos2003,Pace2010,Pace2012,NadkarniGhosh2013,Fan2015,Naderi2015} and clustering dark energy models
\citep{Mota2004,Manera2006,Nunes2006,Abramo2007,Creminelli2010,Basse2011,Batista2013,Malekjani2015,Heneka2017}, 
non-minimally coupled models \citep{Pace2014,NazariPooya2016}, coupled dark energy models 
\citep{Nunes2006,Wintergerst2010a,Tarrant2012}, time varying vacuum cosmologies \citep{Basilakos2009}, decaying dark 
matter with the cosmological constant \citep{Oguri2003} and \cite{Barkana2005,Naoz2005,Naoz2006} included the effects 
of baryons (studying the formation of primordial structures, equations are kept linear).
\footnote{\cite{Basse2011} extended the formalism of clustering dark energy models to arbitrary sound speeds.}  
\cite{Velten2014a} studied the effects of velocity diffusion and \cite{Velten2014c} those of viscosity. 
\cite{Schaefer2008a} used Birkhoff's theorem to derive the equations of motion which were used to study models 
interpolating between the $\Lambda$CDM and the DGP. 
Other works studied the evolution of non-linear perturbations in modified gravity models, such as $f(R)$ 
\citep{Borisov2012,Kopp2013,Lombriser2013a,Cataneo2016}, Galileon \citep{Bellini2012,Barreira2013}, symmetron 
\citep{Taddei2014} and chameleon models \citep{Brax2010,Li2012a,Li2012b}. A parametrised spherical collapse model where 
effects of several screening mechanism are considered is studied in \cite{Lombriser2016}. Finally, 
\cite{Ichiki2012,LoVerde2014} studied the spherical collapse model taking into consideration the effect of neutrinos 
and \cite{Shibusawa2014} the effects of primordial magnetic fields.

In the framework of the spherical collapse model, the collapsing object can be considered as a closed sub-universe 
since in General Relativity Birkhoff's theorem holds. It has spherical symmetry and a uniform density profile 
(top-hat) and being overdense with respect to the background density, it decouples from the Hubble flow, slows down 
and reaches a maximum radius at the turn-around time and starts to collapse. Theoretically the perturbation collapses 
to a point while in reality an equilibrium situation is reached due to the conversion of the energy released during 
the collapsing process into random thermal motion. This is not explicitly present in the formalism and virialization 
has to be inserted a posteriori.

Many improvements to this simple scenario have been introduced: spherical symmetry has been relaxed in favour of an 
ellipsoidal one \citep{Eisenstein1995,Ohta2003,Ohta2004,Angrick2010,Angrick2014,NadkarniGhosh2016}; 
shear, tidal interactions and rotations have also been used to make the collapse more realistic and to evaluate how 
this reflects on the mass function 
\citep{DelPopolo2002,DelPopolo2006b,Cupani2011,DelPopolo2013a,DelPopolo2013b,Pace2014b,Reischke2016a,Reischke2016b,
Pace2017}.

While the top-hat approximation is justified in General Relativity (an initial top-hat profile does not change in 
time), this is not true any more in modified gravity models, such as $f(R)$, where shell crossing develops 
\citep{Borisov2012}.

The main quantity evaluated within this formalism is the linearly extrapolated overdensity $\delta_{\rm c}(z)$ at a 
given collapse redshift $z_{\rm c}$ which represents the value of the linear evolution of the overdensity $\delta$ of 
a perturbation that collapses at the time $z_{\rm c}$ when the full non-linear equation is solved. It is not a quantity 
directly observable, but it represents the main ingredient for the mass function which is, as said above, an important 
tool for cosmology.

When the collapse stops, the radius is formally null and as a direct consequence, the density diverges. As 
\cite{Herrera2017} pointed out recently, a crucial point is how the infinity, or more precisely the numerical infinity 
to describe the collapse, is defined. In this work we aim to extend the analysis of \cite{Herrera2017} and, focusing 
on models within the general relativistic framework, we critically review all the numerical aspects to correctly solve 
the equations of motion of the spherical collapse model. We will concentrate not only on the definition of numerical 
infinity, but also on how to properly define the initial integration time and how to better determine the initial 
conditions.

In \autoref{sect:theory} we review the equations of the spherical collapse model for quintessence models and discuss 
in detail the analytic results for the EdS and the $\Lambda$CDM model, which will be used later on to validate our 
implementation. In \autoref{sect:analysis} we analyse the effects on $\delta_{\rm c}$ of the numerical infinity, of 
the initial time where the integration starts and on the explicit set of the equations adopted, namely a single 
equation for $\delta$ or the combined set of equations for $\delta$ and the divergence of the velocity perturbation 
$\theta$. In \autoref{sect:improvedICs} we detail an improved implementation of the initial conditions and compare our 
numerical findings with the analytic results presented in \autoref{sect:theory}. This comparison will allow us to 
assess the validity and the numerical stability of our implementation for models beyond the $\Lambda$CDM. We want to 
stress that, differently from \cite{Herrera2017}, we do not know in general which value we can expect for 
$\delta_{\rm c}$ for a model with a generic equation of state $w$ or for a $\Lambda$CDM model when the tidal shear and 
rotation invariants are considered. Finally we discuss our findings in \autoref{sect:conclusions} and outline our code 
in more detail in \autoref{sect:code}.

\section{Spherical collapse model}\label{sect:theory}
In this section we review the formalism of the spherical collapse model and introduce its main quantities. We will 
then focus on the analytic results for the EdS and the $\Lambda$CDM model and consider them as a validity test for the 
novel implementation described in detail in \autoref{sect:improvedICs}.

We will follow two different approaches: the first one, based on the evolution of the overdensity $\delta$, has very 
general validity and it will be used to derive the differential equations for $\delta$ which can be applied to a 
generic dark energy model, provided the appropriate background Hubble factor $H(a)$ is used, and to arbitrary 
geometry, once a model for the tidal shear and rotation invariants is assumed; the second one is the original approach 
based on the evolution of the radius of the collapsing sphere and it is very useful to derive analytic results for the 
EdS and the $\Lambda$CDM model.

When using the hydrodynamic approach, we can start from the metric of the perturbed sub-universe
\begin{equation}
 ds^2 = -(1+2\phi)c^2dt^2 + a(t)^2(1-2\phi)\delta_{ij}dx^idx^j\;,
\end{equation}
where $a(t)$ is the scale factor, $\delta_{ij}$ the Kronecker delta and $\phi=\Phi/c^2$ the dimensionless Bardeen 
potential. By assuming the stress-energy tensor of a perfect fluid
\begin{equation}
 T^{\mu\nu} = (\rho+P/c^2)u^{\mu}u^{\nu}+Pg^{\mu\nu}\;,
\end{equation}
where $\rho$ is the density of the fluid, $P = w\rho c^2$ its pressure and $w$ its equation of state. 
The four-velocity is $u^{\mu} = dx^{\mu}/d\tau$, where $\tau$ is the proper time.

Defining the density perturbation $\delta=\delta\rho/\bar{\rho}$ with $\bar{\rho}$ the background density and 
$\boldsymbol{u}$ the three-dimensional comoving peculiar velocity and projecting the continuity equation 
$\nabla_{\mu}T^{\mu\nu}=0$ over the four-velocity $u_{\nu}$ and the projection operator 
$h_{\nu\alpha}=g_{\nu\alpha}+u_{\nu}u_{\alpha}$, we recover the continuity and the Euler equations for the fluid
\begin{align}
 \dot{\delta} + (1+w)(1+\delta)\vec{\nabla}\cdot\boldsymbol{u} = &\; 0\;, \label{eqn:ce}\\
 \dot{\boldsymbol{u}} + 2H\boldsymbol{u} + (\boldsymbol{u}\cdot\vec{\nabla})\boldsymbol{u} + 
 \frac{1}{a^2}\vec{\nabla}\psi = &\; 0\;, \label{eqn:ee}
\end{align}
where $\psi$ is the peculiar gravitational potential which satisfies the following Poisson equation
\begin{equation}\label{eqn:pe}
 \nabla^2\psi = 4\pi G(1+3w)a^2\bar{\rho}\delta\;.
\end{equation}
Spatial derivatives are taken with respect to the comoving coordinate $\boldsymbol{x}$ and a dot represents the 
derivative with respect to cosmic time. Note that in deriving these equations we already have assumed the standard 
"top-hat" profile for the density perturbations.

From now on we consider a pressure-less fluid, namely matter with $w=0$ and density parameter $\Omega_{\rm m}(t)$, and 
we combine the continuity, (the divergence of the) Euler and Poisson equation to obtain the full non-linear 
differential equation describing the evolution of the density perturbation $\delta$ \citep{Ohta2003,Pace2010}:
\begin{equation}\label{eqn:nldelta}
 \ddot{\delta} + 2H\dot{\delta} - \frac{4}{3}\frac{\dot{\delta}^2}{1+\delta} - 
 \frac{3}{2}H^2\Omega_{\rm m}(t)\delta(1+\delta) - (1+\delta)(\sigma^2-\omega^2) = 0\;,
\end{equation}
with $\sigma^2=\sigma_{ij}\sigma^{ij}$ and $\omega^2=\omega_{ij}\omega^{ij}$ the shear and rotation tensors. Under the 
standard top-hat approximation, they arise in the following decomposition
\begin{equation}
 \vec{\nabla}\cdot[(\boldsymbol{u}\cdot\vec{\nabla})\boldsymbol{u}] = \frac{1}{3}\theta^2 + \sigma^2 - \omega^2\;,
\end{equation}
with $\theta=\vec{\nabla}\cdot\boldsymbol{u}$ and are defined as
\begin{align}
 \sigma_{ij} & =\; \frac{1}{2}\left(\frac{\partial u^j}{\partial x^i} + \frac{\partial u^i}{\partial x^j}\right) - 
 \frac{1}{3}\theta\delta_{ij}\;,\\
 \omega_{ij} & =\; \frac{1}{2}\left(\frac{\partial u^j}{\partial x^i} - \frac{\partial u^i}{\partial x^j}\right)\;.
\end{align}

To determine the initial conditions, we seek for an initial overdensity $\delta_{\rm ini}$ such that 
$\delta(a_{\rm c})\rightarrow\infty$, where $a_{\rm c}$ is the collapse scale factor. At early times, $\delta\ll 1$ 
and we can neglect the non-linear terms in Eq.~\ref{eqn:nldelta} and assume a power-law for the linear evolution of 
$\delta$. This procedure will allow us to determine $\dot{\delta}_{\rm ini}$.

Having at hands the initial conditions, we can solve the linearised version of Eq.~\ref{eqn:nldelta}, whose value at 
$a_{\rm c}$ represents the linear overdensity $\delta_{\rm c}$.

It is more convenient to solve the equations of motion having as independent variable the scale factor, so for 
completeness and for clarity in the description of the algorithm, we report both the non-linear and linear differential 
equations:
\begin{align}
 \delta^{\prime\prime} + \left(2+\frac{H^{\prime}}{H}\right)\delta^{\prime} - 
 \frac{4}{3}\frac{{\delta^{\prime}}^2}{1+\delta} - \frac{3}{2}\Omega_{\rm m}(a)\delta(1+\delta) - 
 (1+\delta)(\tilde{\sigma}^2-\tilde{\omega}^2) = 0 \;,\\
 \delta^{\prime\prime} + \left(2+\frac{H^{\prime}}{H}\right)\delta^{\prime}-\frac{3}{2}\Omega_{\rm m}(a)\delta = 0\;,
\end{align}
where a prime represents the derivative with respect to $\ln{a}$ and $\tilde{\sigma}=\sigma/H$ and 
$\tilde{\omega}=\omega/H$. This is what we will refer to as implementation A.

In a more general setup, we might not only have dark matter and baryons clustering, but also dark energy. 
\footnote{Note that since both dark matter and baryons are pressure-less, in this work we consider only their overall 
contribution to the matter component: $\Omega_{\rm m} = \Omega_{\rm cdm} + \Omega_{\rm b}$, where $\Omega_{\rm m}$, 
$\Omega_{\rm cdm}$ and $\Omega_{\rm b}$ are the total matter, cold dark matter and baryon density parameters, 
respectively.} In this case, we need to take into account also possible dark energy perturbations and the equations 
for $\delta_{\rm de}$ will depend on the effective sound speed 
$c_{\rm eff}^2\equiv\delta P_{\rm de}/(\delta\rho_{\rm de}c^2)$. Going from a system of two first order equations 
(one for $\delta$ and one for $\theta=\nabla\cdot\boldsymbol{u}$) to a single second order equation will introduce time 
derivatives of $c_{\rm eff}^2$ and furthermore make the overall equation quite complicated. It is therefore easier, 
also from a numerical point of view since precise initial conditions can be given, to consider the two equations for 
$\delta$ and $\theta$ which, for clustering dark energy models, are 
\citep{Abramo2007,Abramo2008,Abramo2009a,Abramo2009b}
\begin{align}
 \delta^{\prime}_{\rm m}+(1+\delta_{\rm m})\tilde{\theta}_{\rm m} = 0\;,\label{eqn:deltamNL}\\
 \delta_{\rm de}^{\prime}+3\left(c_{\rm eff}^2-w_{\rm de}\right)\delta_{\rm de}+
 \left[1+w_{\rm de}+\left(1+c_{\rm eff}^2\right)\delta_{\rm de}\right]\tilde{\theta}_{\rm de} = 0\;,
 \label{eqn:deltadeNL}\\
 \tilde{\theta}_{\rm m}^{\prime}+\left(2+\frac{H^{\prime}}{H}\right)\tilde{\theta}_{\rm m}+
 \frac{\tilde{\theta}_{\rm m}^2}{3}+(\tilde{\sigma}^2-\tilde{\omega}^2)+
 \frac{3}{2}\left[\Omega_{\rm m}(a)\delta_{\rm m}+\left(1+3c_{\rm eff}^2\right)\Omega_{\rm de}(a)\delta_{\rm de}
 \right] = 0\;,
 \label{eqn:thetamNL}\\
 \tilde{\theta}_{\rm de}^{\prime}+\left(2+\frac{H^{\prime}}{H}\right)\tilde{\theta}_{\rm de}+
 \frac{\tilde{\theta}_{\rm de}^2}{3}+
 \frac{3}{2}\left[\Omega_{\rm m}(a)\delta_{\rm m}+\left(1+3c_{\rm eff}^2\right)\Omega_{\rm de}(a)\delta_{\rm de}
 \right] = 0\;,\label{eqn:thetadeNL}
\end{align}
where we assumed that dark energy is not affected by the $\tilde{\sigma}^2-\tilde{\omega}^2$ term. 
\footnote{Effects of shear and rotation are important only at very late times and for low masses and therefore they do 
not affect in an appreciable way the evolution of dark energy perturbations. Hence their inclusion or exclusion in 
Eq.~\ref{eqn:thetadeNL} will not have any appreciable effect. Including them leads to 
$\tilde{\theta}_{\rm m}=\tilde{\theta}_{\rm de}=\tilde{\theta}$ both at the linear and non-linear level. 
\citep{Pace2014b,Pace2017,Reischke2016b} showed different recipes for the $\sigma^2-\omega^2$ term and that no 
differences are seen when they are (not) included in Eq.~\ref{eqn:thetadeNL}.}
The corresponding linear equations are
\begin{align}
 \delta^{\prime}_{\rm m}+\tilde{\theta}_{\rm m} = 0\;,\label{eqn:deltamL}\\
 \delta_{\rm de}^{\prime}+3\left(c_{\rm eff}^2-w_{\rm de}\right)\delta_{\rm de}+
 (1+w_{\rm de})\tilde{\theta}_{\rm de} = 0\;,\\
 \tilde{\theta}_{\rm m}^{\prime}+\left(2+\frac{H^{\prime}}{H}\right)\tilde{\theta}_{\rm m}+
 \frac{3}{2}\left[\Omega_{\rm m}(a)\delta_{\rm m}+\left(1+3c_{\rm eff}^2\right)\Omega_{\rm de}(a)\delta_{\rm de}
 \right] = 0\;,
 \label{eqn:thetamL}\\
 \tilde{\theta}_{\rm de}^{\prime}+\left(2+\frac{H^{\prime}}{H}\right)\tilde{\theta}_{\rm de}+
 \frac{3}{2}\left[\Omega_{\rm m}(a)\delta_{\rm m}+\left(1+3c_{\rm eff}^2\right)\Omega_{\rm de}(a)\delta_{\rm de}
 \right] = 0\;,
\end{align}
where $\tilde{\theta}=\theta/H$. Therefore at linear level $\tilde{\theta}_{\rm m}=\tilde{\theta}_{\rm de}$. 
This is what we will refer to as implementation B. Note that all the following discussion will be based on the 
assumption that dark energy is important only at the background level. We reported the full equations including dark 
energy perturbations for completeness and we will explain in \autoref{sect:conclusions} what is necessary to be done to 
include them consistently.

Finally, as one may suspect and as \cite{Herrera2017} described at length, a crucial point of the previous approach is 
the definition of numerical infinity, which \cite{Herrera2017} can fix together with the initial conditions for the 
perturbations knowing the result at the collapse time. This is in general not the case when dealing with general dark 
energy models or also for EdS and $\Lambda$CDM models when the assumption of spherical symmetry is dropped, even 
if, in a realistic setting, the difference is at the percent level \citep{Reischke2016a,Reischke2016b}. As shown in 
these works, the problem of infinity can be avoided and a more numerically stable system of equations can be derived 
by noticing that while the density perturbation diverges, its inverse goes to zero. Therefore, defining 
$f\equiv 1/\delta$, our new system is now
\begin{align}
 & f^{\prime} - f(1+f)\tilde{\theta} = 0\;,\label{eqn:fnl}\\
 & \tilde{\theta}^{\prime} + \left(2+\frac{H^{\prime}}{H}\right)\tilde{\theta} + \frac{\tilde{\theta}^2}{3} + 
   \frac{3}{2}\frac{\Omega_{\rm m}(a)}{f} + (\tilde{\sigma}^2-\tilde{\omega}^2) = 0\;,\label{eqn:thetanl}
\end{align}
with corresponding linear equations
\begin{align}
 & f^{\prime} - f^2\tilde{\theta} = 0\;,\label{eqn:fl}\\
 & \tilde{\theta}^{\prime} + \left(2+\frac{H^{\prime}}{H}\right)\tilde{\theta} + 
   \frac{3}{2}\frac{\Omega_{\rm m}(a)}{f} = 0\;.\label{eqn:thetal}
\end{align}
We will refer to this set of equations as implementation C.

To obtain analytic results for the EdS and the $\Lambda$CDM model, it is more convenient to use the formulation which 
follows the evolution of the radius of the sphere. From this point of view, the perturbation, characterised by a given 
constant mass $M$, starts with an initial radius $R_{\rm ini}$ at time $t_{\rm ini}$, evolves to reach a maximum value 
(turn-around radius $R_{\rm ta}$) and decreases to zero when the collapse is reached. In reality this does not happen, 
and the radius will reach a value, called virial radius $R_{\rm vir}$ due to the virialization process. Also in this 
case, virialization needs to be inserted a posteriori in the formalism. Attempts to embed the virialization process by 
modifying the equations of motion have been pursued by \cite{Engineer2000} and \cite{Shaw2008}.

For the derivation of the following results, both at the collapse and at the virialization redshift, we closely follow 
\cite{Lee2010,Lee2010c}. Since explicit analytic results can be obtained for the EdS model, but not in general for a 
$\Lambda$CDM model, here we will only report the EdS ones and refer to the aforementioned works for general 
expressions.

Let us then consider a universe characterised by matter, curvature and dark energy whose equation of state is 
$w_{\rm de}(a)$. Being interested to the late time evolution of the perturbations, we will neglect radiation. 
One can define the scale factor $x$ and radius $y$ of the perturbation normalised at turn-around:
\begin{equation*}
 x \equiv \frac{a}{a_{\rm ta}}\;, \quad y \equiv \frac{R}{R_{\rm ta}}\;,
\end{equation*}
where $R$ is the true radius of the sphere, $a_{\rm ta}$ and $R_{\rm ta}$ the turn-around scale factor and radius, 
respectively.

Denoting by $\omega$ and $\lambda$ the matter and dark energy density parameters at turn-around, respectively, the 
two Friedmann equations describing the collapse are
\begin{align}
 \dot{x} & = \left[\frac{\omega}{x} + \lambda x^2g(x) + (1-\omega-\lambda)\right]^{1/2}\;, \label{eqn:xdot}\\
 \ddot{y} & = -\frac{\omega\zeta_{\rm ta}}{2y^2} - \frac{1+3w(x)}{2}\lambda g(x)y\;,\label{eqn:yddot}
\end{align}
where the dot represents the derivative with respect to the dimensionless time parameter $\tau\equiv H_{\rm ta}t$ and 
$g(x)$ the change in the dark energy density relative to turn-around,
\begin{equation}
 g(x) = \exp{\left\{-3\int_1^x[1+w(x^{\prime})]~{\rm d}\ln{x^{\prime}}\right\}}\;.
\end{equation}
Finally $\zeta_{\rm ta}$ represents the non-linear overdensity of the collapsing sphere with respect to the background 
at the turn-around.

For a constant dark energy equation of state $w_{\rm de}$ and flat spatial geometry, the solution of 
Eq.~\ref{eqn:xdot} is
\begin{equation}
 \tau = \frac{2}{3\sqrt{\omega}}x^{3/2}{}_2F_1\left(\frac{1}{2},-\frac{1}{2w_{\rm de}};1-\frac{1}{2w_{\rm de}};
        -x^{-3w_{\rm de}}\frac{\lambda}{\omega}\right)\;,
\end{equation}
where ${}_2F_1(a,b;c;z)$ is the hypergeometric function. This expression, for $w_{\rm de}=-1$, is identical to what 
was found by \cite{Schaefer2008a}.

For an EdS model, the time of turn-around is therefore $\tau_{\rm ta}=2/3$ ($x_{\rm ta}=1$ by definition and 
${}_2F_1(a,b;c;0)=1$). Since the collapse is symmetric around the turn-around, $t_{\rm c}=2t_{\rm ta}$, we can find the 
relation between the turn-around and the collapse: $a_{\rm ta} = (1/2)^{2/3}a_{\rm c}$.

The solution of Eq.~\ref{eqn:yddot} is quite complicate in general and albeit exact for a $\Lambda$CDM model, it is of 
little use because it can only be inverted numerically. We will therefore concentrate only on a dark energy model 
with $w_{\rm de}=-1/3$ and afterwards take the limit of $\omega=1$. When doing so, we obtain
\begin{equation}
 \zeta_{\rm ta} = \left(\frac{3\pi}{4}\right)^2\left[
                  {}_2F_1\left(\frac{1}{2},\frac{3}{2};\frac{5}{2};-\frac{\lambda}{\omega}\right)\right]^{-2}\;,
\end{equation}
which, for $\omega=1$, gives $\zeta_{\rm ta}=\left(\tfrac{3\pi}{4}\right)^2\simeq 5.55$ \citep{Kihara1968}.

Having this information at hand, we can now evaluate the two most important quantities of the spherical collapse model: 
the linearly extrapolated overdensity at the collapse, $\delta_{\rm c}$, and the virial overdensity $\Delta_{\rm V}$. 
We will also show how to evaluate these same quantities at the virialization redshift. 

For an EdS model, the exact solution of Eq.~\ref{eqn:yddot} is \citep{Lee2010,Lee2010c}
\begin{equation}
 \tau = \frac{1}{\sqrt{\zeta_{\rm ta}}}\left[\frac{1}{2}\arcsin{(2y-1)} - \sqrt{y-y^2} + \frac{\pi}{4}\right]\;,
\end{equation}
which at early times expands into
\begin{equation}
 \tau \approx \frac{8}{9\pi}y^{3/2}\left[1+\frac{3}{10}y\right]\;.
\end{equation}
The overdensity inside the perturbation with respect to the background is
\begin{equation}
 \Delta = \left(\frac{x}{y}\right)^3\zeta_{\rm ta} \approx 1+\frac{3}{5}y\;,
\end{equation}
where the approximation holds at early times. The linear density contrast inside the perturbation is 
$\delta = \Delta - 1$. At turn-around we have $\delta_{\rm ta} = \frac{3}{5}\frac{y}{x}$ and 
$1/x \approx \left(\frac{3\pi}{4}\right)^{2/3}/y$ so that
\begin{equation*}
 \delta_{\rm ta} = \frac{3}{5}\left(\frac{3\pi}{4}\right)^{2/3} \simeq 1.06\;.
\end{equation*}

To determine the virial radius $y_{\rm vir}=R_{\rm vir}/R_{\rm ta}$ one can consider the virialization equation 
combining the classical virial theorem $T = -\frac{R}{2}\frac{\partial U}{\partial R}$, where $U$ is the potential 
energy associated with the overdensity, with the assumption that energy is conserved during the collapse. 
\footnote{We implicitly assumed time averaged quantities.} 
Despite this being a common assumption, for models beyond EdS it has not been proven whether it can actually be 
applied. For example, energy conservation is violated between turn-around and collapse when $w\neq -1, -1/3$ 
\citep{Maor2005,Wang2006,Maor2007}. For an EdS universe, the solution of the equation
\begin{equation}\label{eqn:vir_eq}
 \left[U + \frac{\partial U}{\partial R}\right]_{\rm vir} = U_{\rm ta}\;,
\end{equation}
leads to $y_{\rm vir}=1/2$. For $\Lambda$CDM and dynamical dark energy models a major result is by \cite{Wang1998}:
\begin{equation}
 y_{\rm vir} = \frac{1-\eta_{\rm v}/2}{2+\eta_{\rm t}-3\eta_{\rm v}/2}\;,
\end{equation}
where
\begin{equation*}
 \eta_{\rm t} = 2\zeta_{\rm ta}^{-1}\frac{\Omega_{\Lambda}(a_{\rm ta})}{\Omega_{\rm m}(a_{\rm ta})}\;, \quad 
 \eta_{\rm v} = 2\zeta_{\rm ta}^{-1}\left(\frac{a_{\rm ta}}{a_{\rm vir}}\right)^3
                \frac{\Omega_{\Lambda}(a_{\rm vir})}{\Omega_{\rm m}(a_{\rm vir})}\;.
\end{equation*}
These results can be obtained under the assumption that $y_{\rm vir}$ is only slightly different from the EdS value. 
The algebraic third order equation is then solved perturbatively.

For the virial overdensity, we cannot evolve the full non-linear equation up to the collapse redshift since, by 
definition, $\delta\rightarrow\infty$. We need therefore to use some recipe which allows us to define it. To do so, we 
use the recipe by \cite{Wang1998} and \cite{Maor2005}. 
Several recipes exist in the literature to evaluate the virial overdensity 
\citep{Lahav1991,Kitayama1996,Iliev2001,Battye2003,Weinberg2003,Horellou2005,Maor2005,Wang2006}. 
We performed several tests using all of them and we noticed that they all agree, except for the recipe by 
\cite{Wang2006} (they all agree for an EdS model). This difference was found also in the study of the relativistic 
virialization of structures \citep{Meyer2012}. 
In the following we will limit ourselves to the recipe of \cite{Wang1998} and \cite{Maor2005}.

From the above definition for the non-linear overdensity $\Delta$, and considering the collapse taking place at 
$x_{\rm c}=2^{2/3}$, we find, for an EdS model,
\begin{equation*}
 \Delta_{\rm V}(a_{\rm c}) = 18\pi^2 \simeq 177.65\;.
\end{equation*}
As pointed out by \cite{Lee2010,Lee2010c}, this is not entirely correct, as we should consider $x_{\rm vir}$, rather 
than $x_{\rm c}$. When $y_{\rm vir}=1/2$, $x_{\rm vir}=\left(\frac{3}{2}+\frac{1}{\pi}\right)^{2/3}$ and
\begin{equation*}
 \Delta_{\rm V}(a_{\rm vir}) = 18\pi^2\left(\frac{3}{4}+\frac{1}{2\pi}\right)^2 \simeq 146.84\;.
\end{equation*}

In the limit $\tau\rightarrow 0$, $\Delta = 1+\delta_{\rm lin}$, where \citep{Lee2010,Lee2010c}
\begin{equation}
 \delta_{\rm lin} = \frac{3}{5}\left(\frac{3}{2}\sqrt{\zeta_{\rm ta}}\tau\right)^{2/3}\;.
\end{equation}
At the collapse
\begin{equation*}
 \delta_{\rm c}(a_{\rm c}) = \frac{3}{20}(12\pi)^{2/3} \simeq 1.686\;,
\end{equation*}
while at the virialization
\begin{equation*}
 \delta_{\rm c}(a_{\rm vir}) = \frac{3}{20}(6+9\pi)^{2/3} \simeq 1.58\;.
\end{equation*}
These values will be used to estimate the accuracy of the code.

In the following we will present numerical results only for the linearly evolved overdensity $\delta_{\rm c}$ and show 
results for the virial overdensity $\Delta_{\rm V}$ only for the optimised code, since this quantity is totally 
unaffected by numerical problems or the exact value of the initial conditions. For a more detailed analysis of why 
this is the case, we refer the reader to \cite{Reischke2016a} in the more general analysis of the effect of the shear 
and rotation invariants.

\section{Numerical analysis}\label{sect:analysis}
In the following, we analyse different aspects of the spherical collapse model and show how the final result is 
affected by them. In \autoref{sect:ABC} we compare the three different formulations discussed in \autoref{sect:theory}. 
In \autoref{sect:infinity}, as also done by \cite{Herrera2017}, we show how different definitions of the value for the 
numerical infinity reflect on $\delta_{\rm c}$, while in \autoref{sect:ICs} we discuss the importance of setting 
correctly the initial integration time. Similarly, instead of considering implementations A and B, one could 
consider implementation C and investigate how close to zero the inverse of the density perturbation has to be to 
obtain stable and reliable results. This is discussed in \autoref{sect:zero}.

In this work, we will consider the following cosmological parameters: $\Omega_{\rm m}=0.3$, $\Omega_{\rm de}=0.7$ and 
$h=0.7$. Despite setting initial conditions at very early time, as it is commonly done in the studies of the spherical 
collapse model, we neglect the radiation contribution, setting $\Omega_{\rm r}=0$ throughout the cosmic history.

\subsection{Dependence on the differential equation}\label{sect:ABC}
To solve the three systems of equations, we need to fix some parameters, namely the numerical infinity 
$\delta_{\infty}$, the initial time $a_{\rm ini}$ where the integration starts and the numerical zero 
$f_0=1/\delta_{\infty}$. The differential equations are solved searching for the initial values of $\delta$ 
($\delta^{\prime}$), $\theta$ and $f$ such that the collapse happens at a time specified in input. We consider as 
standard solver a Runge-Kutta Cash-Karp (4, 5) method as implemented in the publicly available library \verb|GSL|.
\footnote{\url{https://www.gnu.org/software/gsl/}} 
We refer the reader to \autoref{sect:improvedICs} for a more detailed discussion about the different choices used to 
evolve the perturbation equations.\\
Here we choose $a_{\rm ini}=10^{-5}$ for all the different implementations, $\delta_{\infty}=10^{8}$ and $f_0=10^{-8}$ 
as a result of the discussion which will follow in the next sections.

\begin{figure}
 \begin{center}
  \includegraphics[scale=0.55,angle=-90]{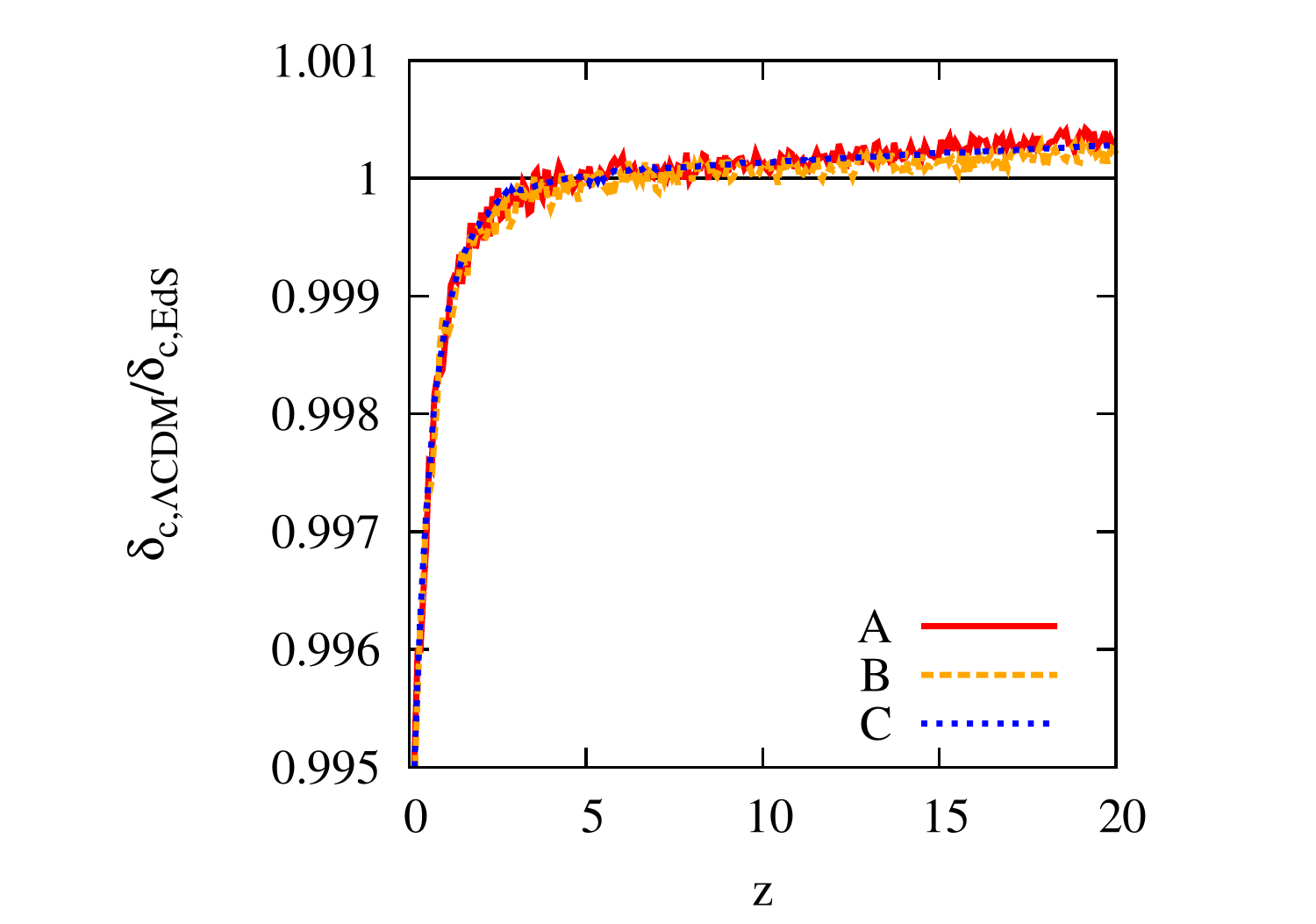}
  \caption{Ratio of $\delta_c$ between $\Lambda$CDM and EdS models as a function of the redshift $z$ for three 
  different implementations. 
  The red solid curve represents the solution of implementation A (a single second order differential equation), the 
  orange dashed curve the solution of implementation B (two first order differential equations, one for $\delta$ and 
  one for $\theta$), while the blue short-dashed curve represents the solution of implementation C (two first order 
  differential equations, one for $f=1/\delta$ and one for $\theta$). The black solid line represents the exact 
  asymptotic value that the ratio should reach.}
  \label{fig:dcABC}
 \end{center}
\end{figure}

In Fig.~\ref{fig:dcABC} we show the evolution of the ratio of $\delta_{\rm c}$ between a $\Lambda$CDM and an EdS model 
for the three different implementations discussed before. As it appears immediately clearly, on average, the solution 
is the same for all of them, as expected. 
In more detail, we see some substantial difference: 
implementations A and B are numerically noisier than implementation C. This is due to the fact that in general, the 
code hits the infinity barrier at slightly different values at each collapse redshift and the initial overdensity 
$\delta_{\rm ini}$ is affected by this. The blue dotted curve instead, is much smoother because it is easier to force 
the code, within a given accuracy, to reach $f_0$. From the point of view of numerical accuracy, implementation C is 
therefore numerically more stable.

Nevertheless, all three solutions show a similar trend which becomes relevant only at high redshifts though 
(differently from \cite{Batista2013} where this was a substantial effect also at lower redshifts): despite approaching 
the EdS solution, $\delta_{\rm c}$ keeps growing and this would have important consequences on the mass function of 
high redshifts objects, with the result of predicting fewer objects. We also point out that at such high redshifts 
($z\gtrsim 8$) an appropriate mass function must be used \citep[see e.g.][for an in-depth discussion]{DelPopolo2017}.

This is due to the fact that moving towards higher redshifts, $\delta_{\rm ini}$ grows too fast with respect to what 
is required (see \autoref{fig:ini}), because of small imprecisions arising from the combined effect of the root-finding 
algorithm, of the effective value of collapse $\delta_{\infty}$ ($1/\delta_{\infty}$) and of how the initial conditions 
are defined. 
Either solving a second order differential equation or a system of two differential equations, we require two initial 
conditions: for implementation A we need to find $\delta_{\rm ini}$ and $\delta_{\rm ini}^{\prime}$ while for 
implementation B we need $\delta_{\rm ini}$ and $\theta_{\rm ini}$ and for implementation C $f_{\rm ini}$ and 
$\theta_{\rm ini}$. Once $\delta_{\rm ini}$ or $f_{\rm ini}$ are found, $\delta_{\rm ini}^{\prime}$ and 
$\theta_{\rm ini}$ are related to them either via differentiation or via the continuity equation. Hence it is clear 
that a not exquisitely accurate determination of $\delta_{\rm ini}$ will have appreciable consequences.

Note however that the difference from a constant value is at the sub-percent level and of the order of 1-2\% at most 
for $z\simeq 40$, making therefore results in literature generally correct.

As we will explain later in detail, this side effect can be avoided to a very high degree by refining the 
determination of the initial conditions.

\subsection{Dependence on the collapse value}\label{sect:infinity}
\cite{Herrera2017} pointed out clearly that results for $\delta_{\rm c}$ for a $f(R)$ model depend on the chosen value 
of the numerical infinity adopted and it is not correct to keep it constant when exploring the collapse at higher 
redshifts. The authors considered two different values for the collapse today: $\widetilde{\rm Inf}=10^5$ and 
$\widetilde{\rm Inf}=10^8$ setting the initial conditions at $z_{\rm ini}=1000$ and showed that this value can be as 
low as $10^4$ at collapse redshift $z\sim 3$ and that differences between the standard method and the method presented 
by the authors (based on the fact that they know what $\delta_{\rm c}$ should be at $z_{\rm c}=0$) grow increasing the 
redshift, since $\delta_{\rm c}$ grows without bound, as clearly shown in Fig.~\ref{fig:dcABC}.

\begin{figure}
 \begin{center}
  \includegraphics[scale=0.55,angle=-90]{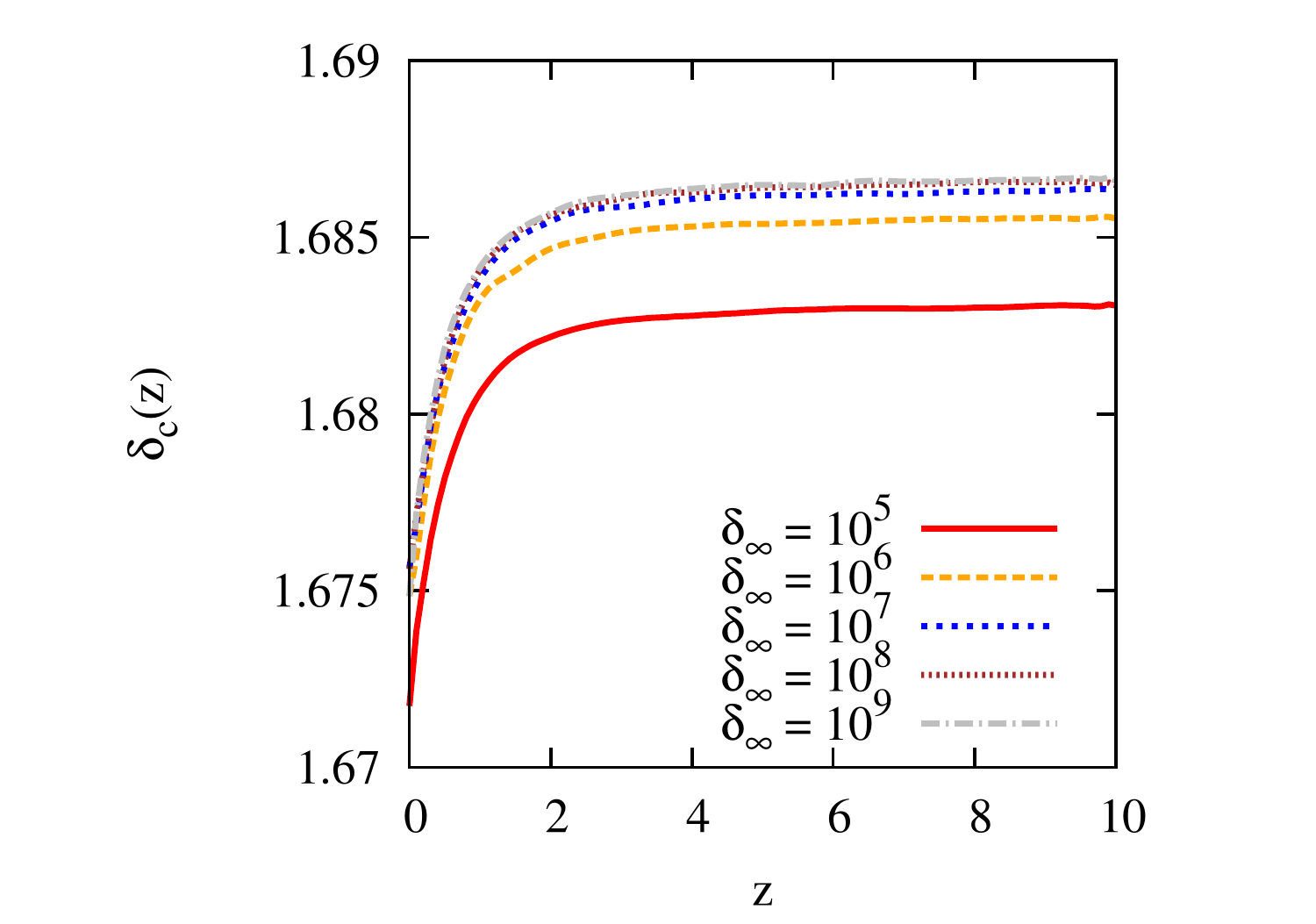}
  \caption{Critical collapse density $\delta_{\rm c}$ as a function of $z$ for different values of $\delta_{\infty}$. 
  From bottom to top we consider: $10^5$ (red solid line), $10^6$ (orange dashed line), $10^7$ (blue short-dashed 
  line), $10^8$ (brown dotted curve) and $10^9$ (grey dot-dashed curve), respectively.}
  \label{fig:dcInfinity}
 \end{center}
\end{figure}

We want to investigate further the dependence of $\delta_{\rm c}$ on the exact value of the numerical infinity. The 
question we want to answer is: when can we be confident that the code converged, i.e., that when increasing the 
numerical infinity, the result does not change appreciably? We show this in Fig.~\ref{fig:dcInfinity}, where we 
considered values of $\delta_{\infty}$ ranging from $10^{5}$ to $10^{9}$ for implementation B. Note also that this 
value is fixed and it does not change at higher collapse redshifts.

It is clear that $\delta_{\rm c}$ depends critically on $\delta_{\infty}$. For a value as low as 
$\delta_{\infty}=10^5$, the value for $\delta_{\rm c}$ is substantially lower than the value expected, with a 
difference of about 2\textperthousand. This is roughly in agreement with what was found by \cite{Herrera2017}. 
Note that even if the difference might seem irrelevant and probably it is now with current data, it is not at all 
marginal for the mass function and the halo number counts for precision studies aiming to infer dark energy properties. 
Since at high mass, where we expect dark energy and modified gravity to play a considerable role, the mass function 
depends exponentially on the value of $\delta_{\rm c}$ and sub-percent differences here translate in differences of 
several percent in the mass function and quantities based on it, such as the Sunyaev-Zel'dovich (SZ) power spectrum. 
The evolution of $\delta_{\rm c}$ depends also on the physics considered, such as shear and rotation, or different 
gravitational models and it is therefore essential to have a converged result.

Increasing $\delta_{\infty}$ results in an increase of $\delta_{\rm c}$, such that for values of 
$\delta_{\infty}\geq 10^7$, the numerical solution has essentially converged to the expected value. Note that in 
Fig.~\ref{fig:dcInfinity} it is still possible to appreciate differences between $\delta_{\infty}=10^7$ and 
$\delta_{\infty}=10^8$, while for higher values differences are less than one part in $10^4$, approximately constant 
over all the redshifts investigated. We will hence consider from now on $\delta_{\infty}=10^8$.

Note that here, for clarity purposes, we smoothed the curves to suppress the numerical noise, so to be able to 
distinguish results for $\delta_{\infty}\geq 10^7$.

\subsection{Dependence on the Time of the Initial Conditions}\label{sect:ICs}
Another important quantity deserving attention is the time $a_{\rm ini}$ (or equivalently $z_{\rm ini}$), when initial 
conditions are set. Setting them too early will unnecessarily slow down the code, unless adaptive step-sizes are used, 
since we expect an EdS behaviour at high redshifts, unless early dark energy models are considered 
\citep{Doran2001,Wetterich2004,Doran2005b,Doran2006,Bartelmann2006,Francis2009a,Francis2009b,Pettorino2013}. 
On the other hand, setting them too late might not lead the code to the expected solution. This is because formally 
perturbations originate when $a_{\rm ini}\simeq 0$.

Here we recall how initial conditions are determined. The particular implementation is totally irrelevant, so we will 
just outline the general procedure. For a given collapse time $a_{\rm c}$, we need to infer the value of the initial 
overdensity $\delta_{\rm ini}$ such that $\delta(a_{\rm c})\rightarrow\infty$ or $f(a_{\rm c})\rightarrow 0$. In 
principle, $\delta_{\rm ini}$ can be specified as an input parameter, but this will lead to a collapse time that in 
general does not coincide with the desired one. The differential equation for $\delta_{\rm c}$ is then evolved having 
$\delta_{\rm ini}$ as initial condition.

\begin{figure}
 \begin{center}
  \includegraphics[scale=0.55,angle=-90]{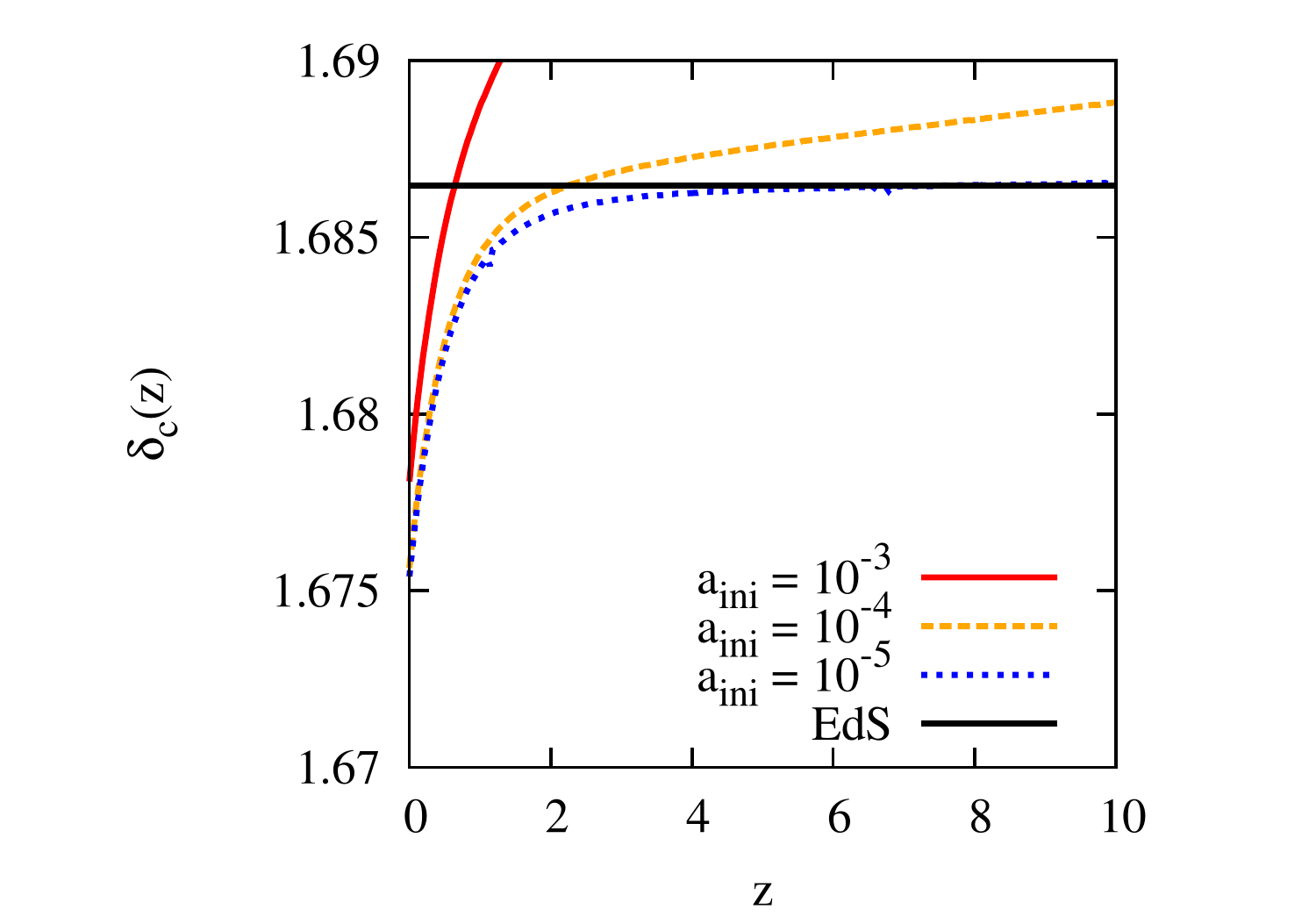}
  \caption{Evolution of $\delta_{\rm c}$ as a function of $z$ for a $\Lambda$CDM model for different values of 
  $a_{\rm ini}$. The horizontal black solid curve represents the exact value of an EdS universe, the red solid curve 
  the solution for $a_{\rm ini}=10^{-3}$, the orange (blue) dashed (short-dashed) curve the value for 
  $a_{\rm ini}=10^{-4}$ ($a_{\rm ini}=10^{-5}$), respectively.}
  \label{fig:dcAi}
 \end{center}
\end{figure}

In Fig.~\ref{fig:dcAi} we show the influence of the choice of $a_{\rm ini}$ on the evolution of the linearly 
extrapolated overdensity $\delta_{\rm c}$, as a function of redshift. Once again we consider a $\Lambda$CDM model. A 
similar behaviour happens in an EdS model. We present results for implementation C to have a smoother outcome, 
but we checked that a very similar result is found when considering the equations for $\delta$.

It is clear that to achieve a reliable result requires also an early time to start integrating the differential 
equations. We investigated different cases, ranging from $a_{\rm ini}=10^{-3}$ to $a_{\rm ini}=10^{-5}$. For large 
values of $a_{\rm ini}$, the numerical solution has certainly not converged since the perturbation did not collapse yet 
and as a consequence $\delta_{\rm c}$ grows unboundedly. We can observe a clear flattening of the curves decreasing the 
initial integration time. For $a_{\rm ini}$ as small as $10^{-5}$, the solution has definitely converged to the 
expected value (shown by the black solid line for the EdS model).

It is worth to investigate this aspect more in detail. \cite{Herrera2017} set their initial conditions at 
$z_{\rm ini}=10^3$ and find a reasonable solution for $\delta_{\rm c}$, for both $\delta_{\infty}=10^5$ and $10^8$, 
albeit not converging to the expected EdS solution (it grows unboundedly). \cite{Batista2013}, in the context of 
clustering early dark energy models, showed the evolution of $\delta_{\rm c}$ for a $\Lambda$CDM model. 
It converges to the numerical solution for an EdS cosmology for $z\simeq 4-5$, but both of them do not flatten to the 
expected analytic value. Nevertheless, their result is, within at most few percent, in agreement with what one has to 
expect. 
When instead we fix the value of $\delta$ (its inverse) to high (small) values, we find indeed an acceptable solution 
only for small enough $a_{\rm ini}$.

This result can be explained taking into account the settings of the different codes and the results of 
Fig.~\ref{fig:dcInfinity}. We keep both $\delta_{\infty}$ and $a_{\rm ini}$ fixed and we do not change the first as a 
function of the collapsing time. The initial overdensity increases monotonically increasing the collapse redshift, but 
we do not have any scaling of our initial density perturbations since, for a generic dark energy model, we do 
not know a priori what the value for $\delta_{\rm c}$ should be, even if for a realistic model, it can not be too 
different from the $\Lambda$CDM value. This explains our findings with respect to \cite{Herrera2017}. When instead 
comparing our implementation with \cite{Batista2013}, we can appreciate the interplay and the degeneracy of 
$\delta_{\infty}$ and $a_{\rm ini}$. The authors fix the initial conditions at $z_{\rm ini}=10^3$ and assume 
$\delta_{\infty}=10^5$. According to our analysis, the first condition gives a too high value for $\delta_{\rm c}$, 
but a too low value for the latter. Both effects partially cancel each other delivering a good value for the 
linearly extrapolated overdensity. At the same time we notice that the sharp rise of $\delta_{\rm c}$ (our red solid 
line in Fig.~\ref{fig:dcAi}) is only partially mitigated by a low value of $\delta_{\infty}$ in Fig.~7 of 
\cite{Batista2013}, being somehow intermediate with respect to our curves for $a_{\rm ini}=10^{-3}$ and 
$a_{\rm ini}=10^{-4}$.

This is a strong indication that not only the initial conditions have to be determined exquisitely well, but also all 
the other parameters can play a very important role if an analytic solution is not known. We will return to this 
point in \autoref{sect:improvedICs}, where we discuss an alternative way to evaluate the initial conditions by scaling 
$a_{\rm ini}$.

\subsection{Dependence on the extrapolation}\label{sect:zero}
In Fig.~\ref{fig:dcABC} we showed that evolving the equation describing the evolution of $f=1/\delta$ gives a much 
less noisy result. While for the evolution of $\delta$ is critical to use a high enough value of $\delta_{\infty}$ for 
a proper numerical convergence to the analytic result, for $1/\delta$ it is important to know how close one has to be 
to zero to obtain the right answer. This is what we will discuss in this section. In the following we will consider, 
following the discussion in the previous section, $a_{\rm ini}=10^{-5}$.

While numerically it is difficult to compare two big numbers, it is much easier to compare two small numbers within a 
given precision range. Evolving $f=1/\delta$ implies as collapse condition that $f\rightarrow 0$. This cannot be 
achieved exactly but we can enforce it in this way: we evolve the system of differential equations till the collapse 
scale factor and till $f>\varepsilon=f_0$, but in general none of this conditions will be achieved exactly. The solver 
returns the value of $f$ and its derivative and we use this information to linearly extrapolate the value of $f$ to 
the exact collapse scale factor. In this way our solution is much less affected by the numerical noise due to the fact 
that the collapse condition and the collapse time are now reached exactly (within machine precision).

\begin{figure}
 \begin{center}
  \includegraphics[scale=0.55,angle=-90]{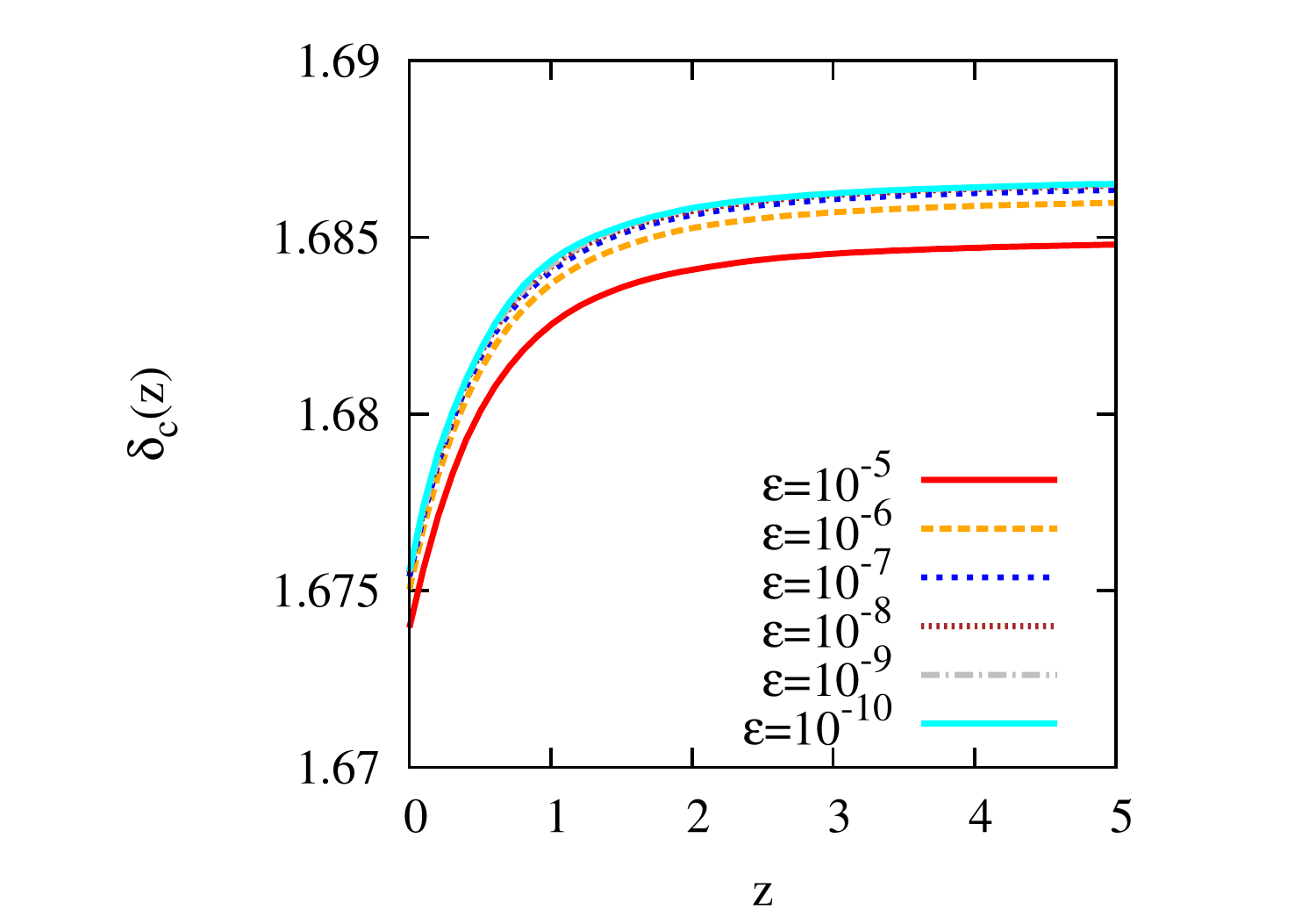}
  \caption{Evolution of $\delta_{\rm c}$ as a function of $z$ for a $\Lambda$CDM universe for different values of the 
  extrapolation parameter $\varepsilon$. From bottom to top: the red (cyan) solid curve represents the solution for 
  $\varepsilon=10^{-5}$ ($\varepsilon=10^{-10}$), the orange dashed curve $\varepsilon=10^{-6}$, the blue 
  short-dashed curve $\varepsilon=10^{-7}$ and the brown dotted (grey dot-dashed) the solution for 
  $\varepsilon=10^{-8}$ ($\varepsilon=10^{-9}$), respectively.}
  \label{fig:dcZero}
 \end{center}
\end{figure}

It is therefore important to determine how small $\varepsilon$ must be to achieve convergence of the result. We show 
the results of our analysis in Fig.~\ref{fig:dcZero}.

The situation is very similar to what was found and discussed before for $\delta_{\infty}$. When $\varepsilon=10^{-5}$, 
the perturbation is not collapsed yet and $\delta_{\rm c}$ is smaller than what it should be. This is in agreement with 
the results presented in Fig.~\ref{fig:dcInfinity}: when $\delta_{\infty}=10^5$, the perturbation has not collapsed yet 
and the numerical value of $\delta_{\rm c}$ is smaller than the true one. The main difference is that the error now is 
smaller than before. Decreasing the value of $\varepsilon$ brings the numerical solution closer to the true one, so 
that, as expected from the discussion above, for $\varepsilon\leqslant 10^{-8}$, the numerical solution has completely 
converged to the true value. In the next section we will therefore assume $\varepsilon=10^{-8}$.

\section{Improved Initial Conditions}\label{sect:improvedICs}
After the discussion in \autoref{sect:analysis}, we can now use the main results obtained to implement the equations 
of motion describing the evolution of the perturbations in an improved code. After discussing the details of the 
determination of the improved initial conditions and the structure of the code, we compare our numerical results with 
the analytic ones. 
According to the previous analysis, in the following we will consider the following parameters: $\delta_{\infty}=10^8$, 
$\varepsilon=10^{-8}$ and the maximum value to start integrating the equations $a_{\rm ini}=10^{-5}$.

As already discussed in detail in \autoref{sect:analysis}, fixed initial conditions as well as a fixed value of 
numerical infinity lead to an artificial dependence of $\delta_{\rm c}$ on the chosen value of the collapse redshift. 
In the EdS limit, this leads to a small linear increase of the overdensity $\delta_{\rm c}$ with collapse redshift. 
\cite{Herrera2017} proposed a method to obtain the required value of the numerical infinity $\widetilde{\mathrm{Inf}}$ 
as a function of $z_{\rm c}$. This method requires calibrating the initial density contrast using the analytical 
results for the linear overdensity $\delta_{\rm c}$ in the EdS model where $\widetilde{\mathrm{Inf}}$ is assumed to be 
roughly independent of the cosmological model. We decided to use an alternative approach based on the appropriate 
rescaling of the initial scale factor to approximately match a constant value of the numerical infinity. 
As discussed in the previous section, the nonlinear evolution is described by the reciprocal density contrast $f$ and 
we therefore fix a lower bound $\epsilon$ to this quantity. The initial density contrast is then obtained by 
extrapolating $f_{\rm nl}$ to zero leading to the given initial value $f_{\rm ini} = 1/\delta_{\rm ini}$. 
Since this procedure generically causes the value for the numerical infinity to depend on the collapse redshift, we 
need to change the initial conditions or equivalently the range of integration of the differential equation system to 
compensate for that. It turns out that an appropriate scaling for the initial scale factor is given by the linear 
growth factor of the model evaluated at the collapse scale factor. We therefore apply a simple rescaling law
\begin{equation*}
 a_{\rm ini} \longrightarrow \frac{a_{\rm ini}}{D_{+}(a_{\rm c})}\;.
\end{equation*}

By truncating the range of integration at a certain $a_{\rm c}<1$, the logarithmic range of numerical integration 
decreases by the amount $\sim -\log(D_+(a_{\rm c}))$ once the initial scale factor is fixed. Using this rescaling law, 
we keep the logarithmic size of integration interval approximately constant in $\log(a)$. 
Once the initial scale factor is rescaled, a constant value of numerical infinity is approximately matched.
\footnote{We checked that for an EdS model, we find indeed a constant value for the numerical infinity 
($\widetilde{\rm Inf}=1.00372\times 10^{8}$) and $10^{-19}<|\epsilon|<10^{-17}$, indeed very close to zero. For a 
$\Lambda$CDM model we find the same range for $|\epsilon|$ and small variations for $\widetilde{\rm Inf}$: in 
particular, between $z=0$ and $z=20$, we have 
$(\widetilde{\rm Inf}_{\rm max}-\widetilde{\rm Inf}_{\rm min})/\widetilde{\rm Inf}_{\rm min}\approx 10\%$. This shows 
once again that an exact determination of the initial conditions is more important that the exact value of the 
numerical infinity, provided this is high enough to have the solution numerically converged.} 
In case of the EdS model, the time evolution of $f$ is independent of the scale factor $a_{\rm c}$ and therefore a 
constant value of numerical infinity is matched exactly. In case of more complicated background fluids, this holds only 
approximately as the functional form of $f(a)$ changes with collapse redshift. We therefore rescale the initial scale 
factor by $D_+(a_{\rm c})$ instead of $a_{\rm c}$. Despite of that, this method does not require any calibration from 
an analytical result. Once this rescaling of the initial conditions is applied, results are perfectly stable up to 
collapse redshift of $20$ (the maximum we investigated) and reproduce the correct EdS value for $\delta_{\rm c}$ 
without initial calibration.

The following step is a further polishing of the initial conditions via a Newton-Raphson method. For the sake of 
computation time, we limit the code to five iterations. Starting from the initial guess of $\delta_{\rm ini}$, we 
evaluate the nonlinear value of $f$ and $\partial f/\partial\delta_{\rm ini}$ at the collapse and we move 
$\delta_{\rm ini}$ along the slope of the curve given by $f/(\partial f/\partial\delta_{\rm ini})$. If the change is 
smaller than about 10\%, we stop the refinement.

\begin{figure}
 \begin{center}
  \includegraphics[scale=0.55,angle=-90]{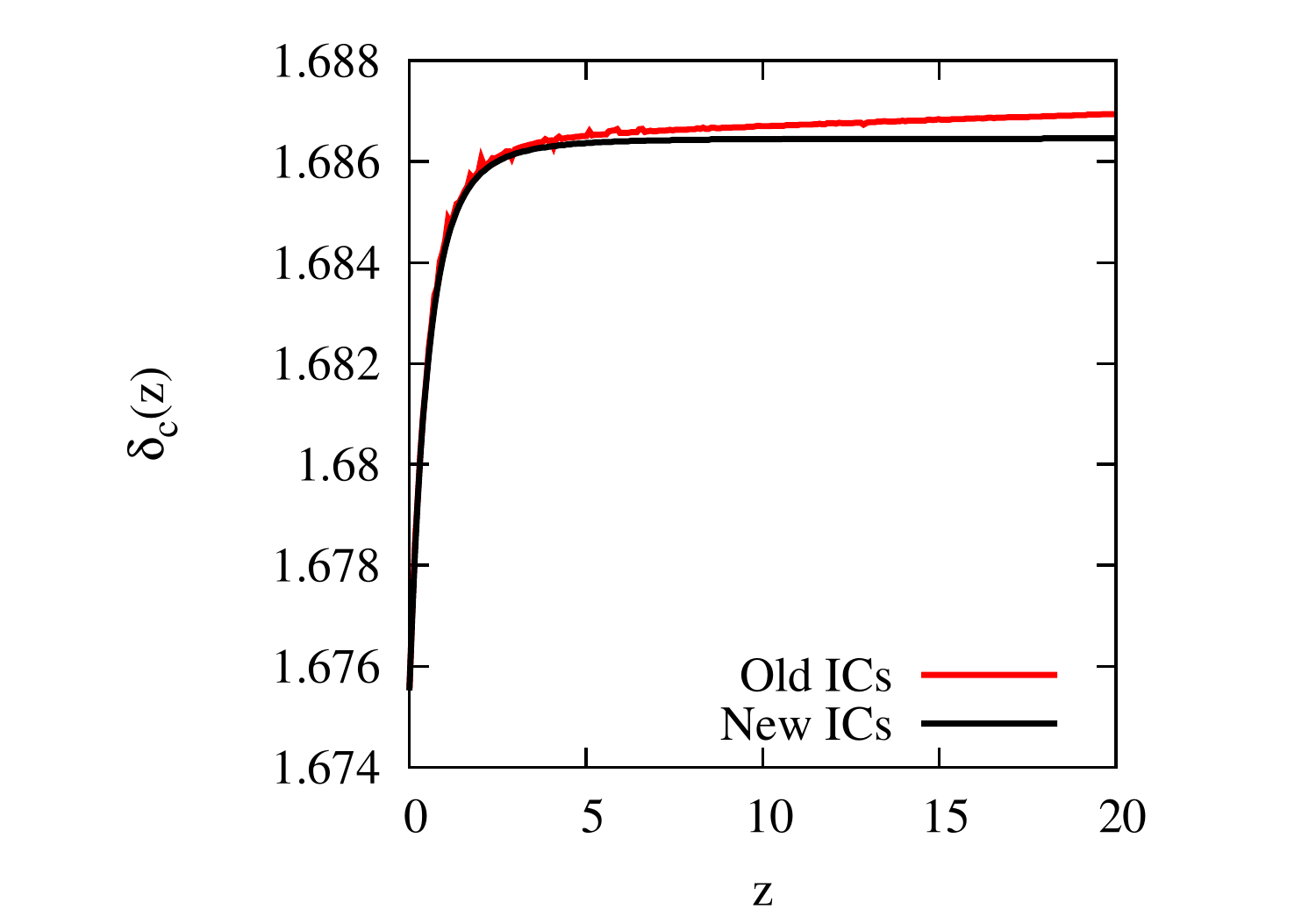}
  \caption{Comparison of $\delta_{\rm c}$ for different choices of initial conditions. The black (red noisy) solid 
  curve represents the $\Lambda$CDM model with the novel (old) determination of the initial conditions. In both cases 
  we used implementation C.}
  \label{fig:deltacNew}
 \end{center}
\end{figure}

In Fig.~\ref{fig:deltacNew} we show the evolution of $\delta_{\rm c}$ for the two different initial conditions for a 
$\Lambda$CDM model. We consider implementation C for the old and new determination of the initial conditions. It is 
immediately clear that the improved initial conditions greatly improve the result. First of all the residual numerical 
noise due to the not exact estimation of the initial conditions is removed (for implementation A the numerical noise 
will be higher, but the time dependence identical). The initial overdensity is now accurate at the machine level thanks 
to the polishing procedure. In addition, and more importantly, $\delta_{\rm c}$ is now bounded: it reaches the 
theoretical value for an EdS model exactly and does not grow indefinitely, differently from before. This is a very 
important achievement showing the power of the new implementation. As a practical consequence, numerical values for 
$\delta_{\rm c}$ can now be used for the study of objects at high redshifts also with the help of the mass function, 
as discussed by \cite{Watson2013}.

Having established the reliability of our code, we can now study in more detail some other quantities characterising 
the spherical collapse model. Apart from the value of $\delta_{\rm c}$ at the collapse, we can also study its evolution 
at the virialization redshift. While $\delta_{\rm c}$ represents the linear evolution of $\delta$, we can also 
investigate its non linear evolution, namely the virial overdensity $\Delta_{\rm V}$ at both the collapse and the 
virialization redshifts.

Since the collapse proceeds at a faster pace approaching the collapse time and the divergence appears formally only at 
the collapse, it is safe to integrate the full non-linear equation for $\delta$ till the virialization time. 
We verified that doing this is in excellent agreement with applying the recipe of \cite{Wang1998} replacing all the 
quantities evaluated at collapse with the corresponding ones at the virialization.

To evaluate the turn-around scale factor $a_{\rm ta}$, we solve Eqs.~\ref{eqn:fnl} and \ref{eqn:thetanl} and 
determine $\log[(\delta_{\rm nl}+1)/a^3]$, which, besides irrelevant constants, is the inverse of the sphere radius 
$y$. The maximum value of $y$ is reached at the turn-around, therefore its inverse will reach at the same time its 
minimum. Hence, finding the location of the minimum will automatically give the value of the turn-around. Integrating 
the same set of equations till the turn-around will return $\zeta_{\rm ta}-1$.

Finally, the last quantity to evaluate is the virial scale factor $a_{\rm v}$. To do so, we first choose a recipe for 
the virialization process among the works cited above which provides the quantity $y_{\rm vir}$, then we look for the 
scale factor $x_{\rm v}=a_{\rm v}/a_{\rm ta}$ which minimises the function 
$\sqrt[3]{\zeta_{\rm ta}/(1+\delta_{\rm nl})}x_{\rm v}-r$. When this function is null with a precision of $10^{-8}$, 
the code returns the guessed value of $a_{\rm v}$. We repeat this procedure till the estimated virial scale factor does 
not change more than one part in $10^8$.

Later on we will compare the turn-around time $a_{\rm ta}$ and the non-linear overdensity $\zeta_{\rm ta}$ with 
theoretical predictions.

\begin{figure}
 \begin{center}
  \includegraphics[scale=0.5,angle=-90]{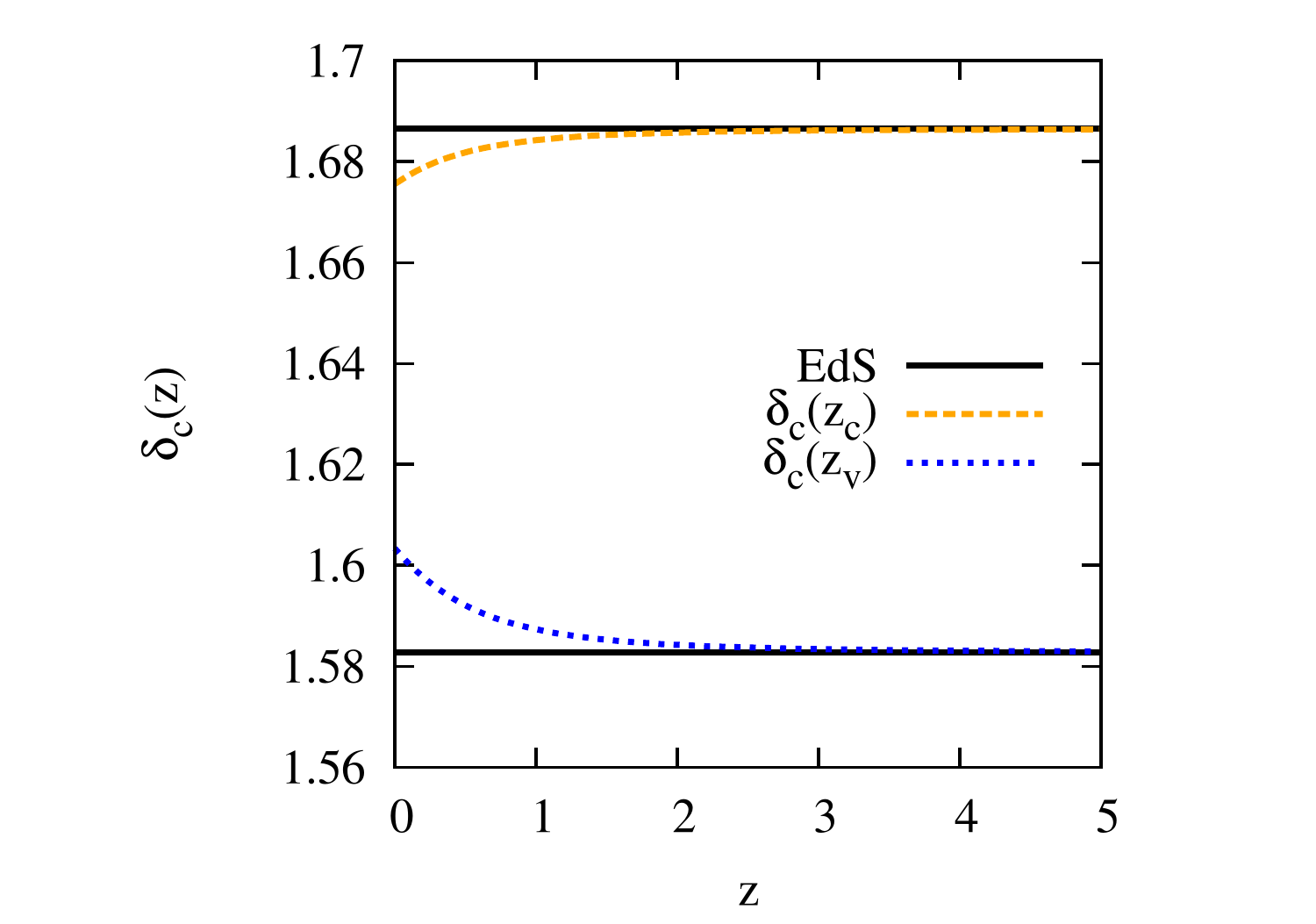}
  \includegraphics[scale=0.5,angle=-90]{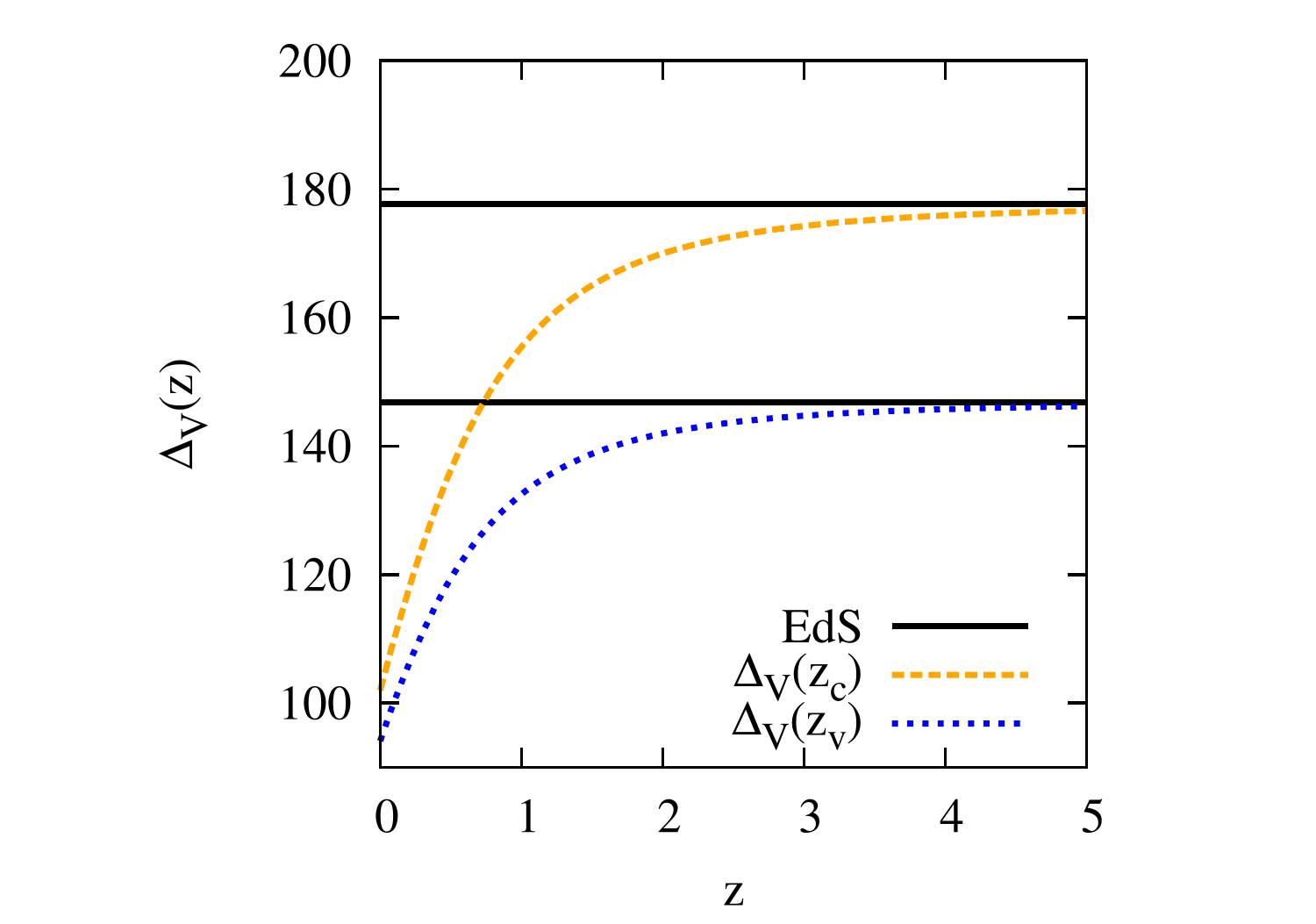}
  \caption{Left (right) panel: Evolution of $\delta_{\rm c}$ ($\Delta_{\rm V}$) evaluated at collapse $z_{\rm c}$ and 
  virialization $z_{\rm v}$ redshift, respectively. The two horizontal black solid lines represent the EdS model, 
  while the orange dashed and blue short-dashed lines the $\Lambda$CDM model. Upper (lower) curves show quantities at 
  collapse (virialization) redshift.}
  \label{fig:deltav}
 \end{center}
\end{figure}

In Fig.~\ref{fig:deltav} we show the time evolution of $\delta_{\rm c}$ (left panel) and 
$\Delta_{\rm V}=\zeta_{\rm ta}\Omega_{\rm m}(a_{\rm c,v})(x/y)^3$ (right panel) as a function of the collapse 
($a_{\rm c}$) or virialization ($a_{\rm v}$) scale factor for the EdS and $\Lambda$CDM models. It is clear that now 
the linear overdensity reaches exactly the theoretical value of an EdS model and does not grow further. This holds for 
both the EdS and the $\Lambda$CDM model, either at collapse or virialization. While the right panel does not reveal any 
surprise (the virial overdensity is smaller at late times where the cosmological constant dominates and approaches the 
EdS value for $z\gtrsim 5$), the left panel presents an interesting result. The upper curves show the linear 
overdensity at the collapse: at early times the two models give the same results as the EdS model is an excellent 
approximation, but at low redshift we have a different situation. When we consider the collapse time, the linear 
overdensity for the $\Lambda$CDM model is smaller than for an EdS model: this can be easily explained by taking into 
account that it is necessary to have a lower threshold for collapse because of the presence of the cosmological 
constant (analogous results are for generic dark energy models). When instead we consider the time of virialization, 
the expected value for $\delta_{\rm c}$ is higher for a $\Lambda$CDM than for an EdS one, while naively one could have 
thought the opposite behaviour, in analogy to the collapse. To explain this, it is useful to recall that 
$\delta_{\rm c}$ enters in the determination of the mass function: the number of massive objects must be the same both 
at the time of collapse and of virialization, since we consider a collapsed object virialized and vice versa. Hence one 
should not focus on $\delta_{\rm c}$ itself, but on the quantity 
$\nu_{\rm c}=\delta_{\rm c}(a_{\rm c})/D_{+}(a_{\rm c})$ which enters into the mass function. In the definition of 
$\nu_{\rm c}$, $D_{+}$ represents the linear growth factor. We checked that $\nu_{\rm c}=\nu_{\rm v}$ thus explaining 
the result in the left panel of Fig.~\ref{fig:deltav}. In other words, $\nu$ is a measure of $\delta_{\rm ini}$.
We also compared our numerical solution with the fit provided by \cite{Schaefer2008a} for a flat $\Lambda$CDM model 
with $\Omega_{\rm m}=0.25$. For $0<z<5$, we found an agreement of the order 0.2\%-0.3\%.

\begin{figure}
 \begin{center}
  \includegraphics[scale=0.5,angle=-90]{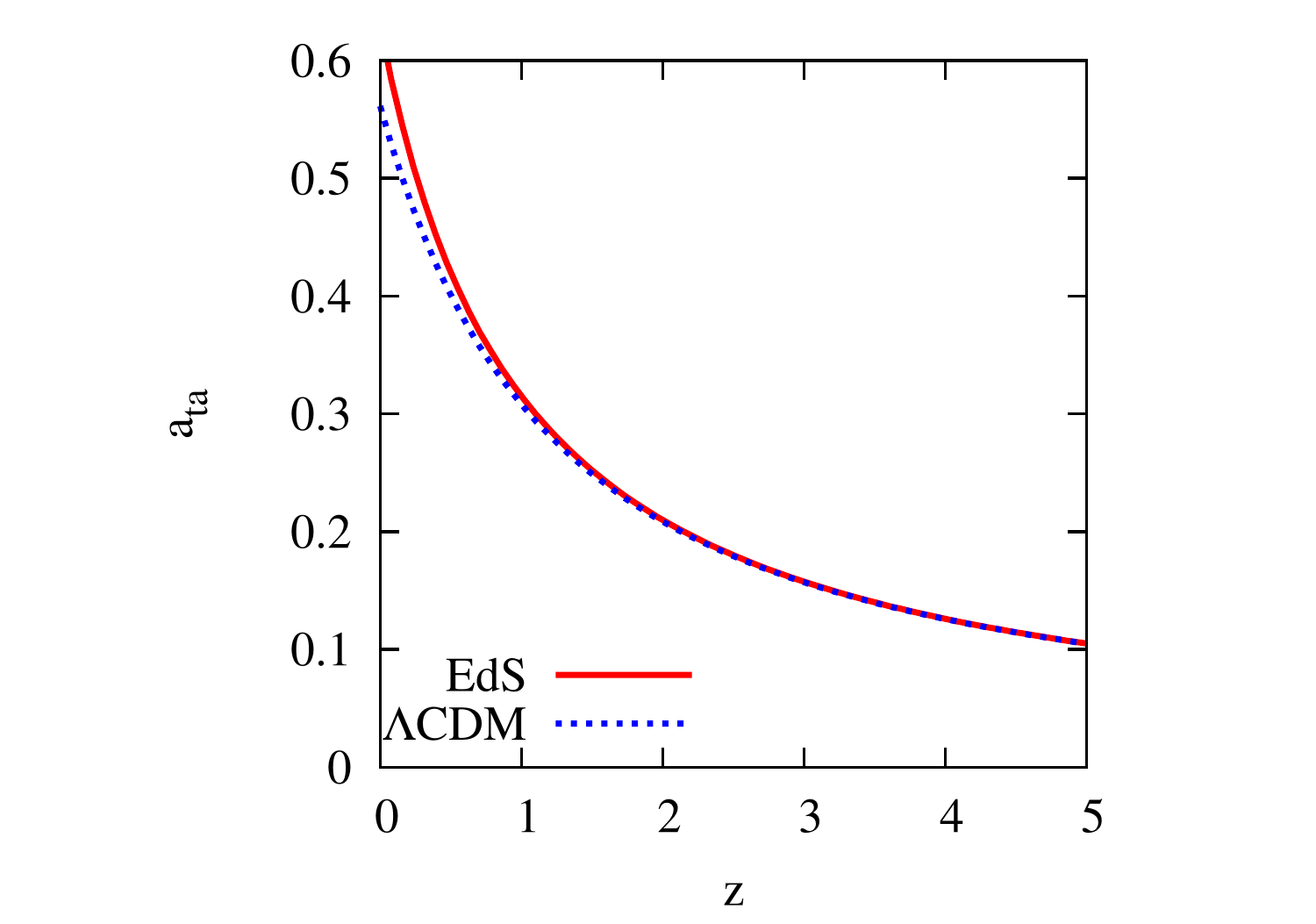}
  \includegraphics[scale=0.5,angle=-90]{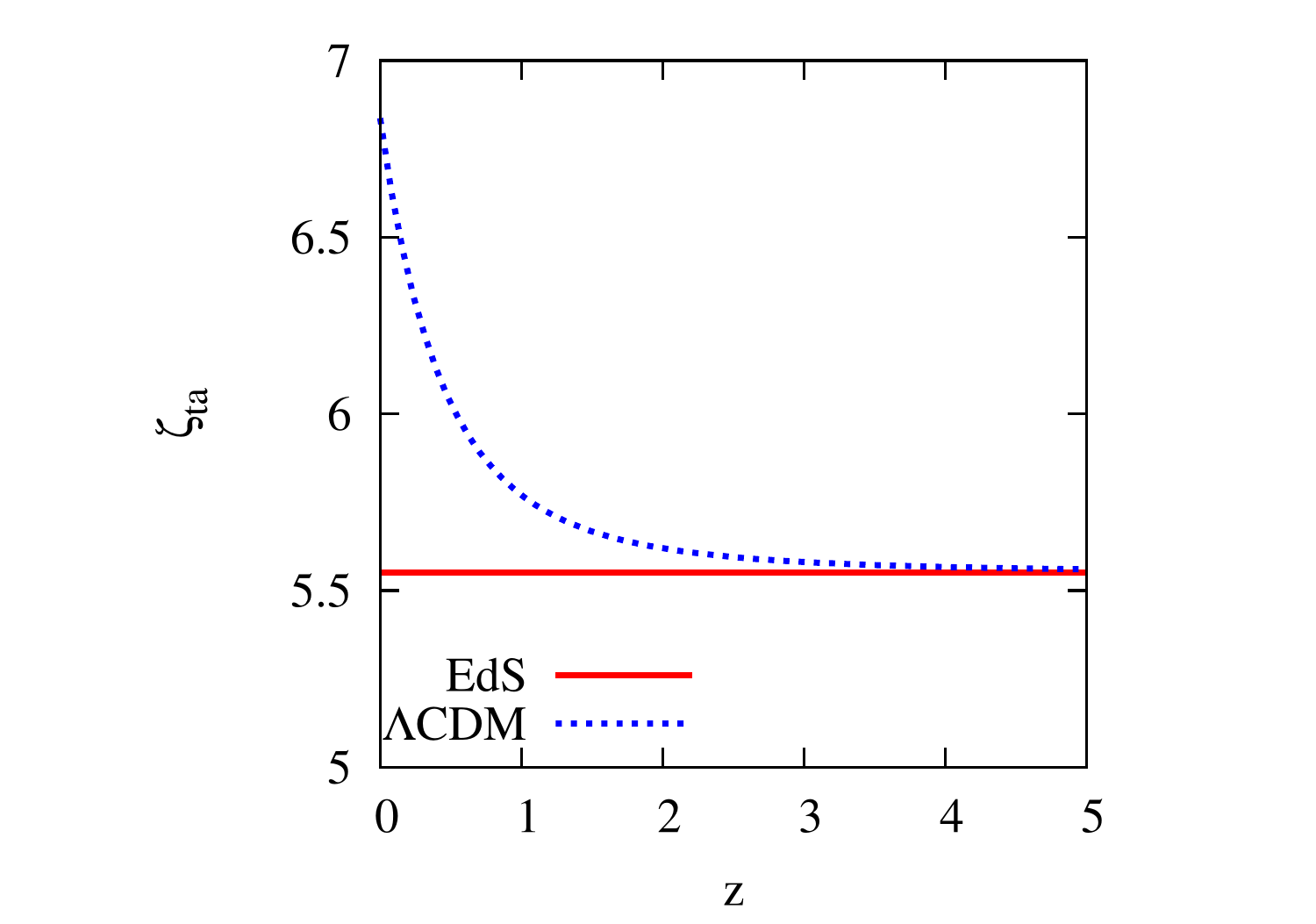}
  \caption{Left panel: evolution of the turn-around scale factor $a_{\rm ta}$ as a function of the collapse redshift. 
  Right panel: evolution of the non-linear density $\zeta_{\rm ta}$ at turn-around. The solid red line shows the result 
  for the EdS model, the dashed blue line for the $\Lambda$CDM result.}
  \label{fig:ata_zeta}
 \end{center}
\end{figure}

In Fig.~\ref{fig:ata_zeta} we show the turn-around (left panel) and the corresponding value of the density 
$\zeta_{\rm ta}$ (right panel), again for an EdS and a $\Lambda$CDM model. Remember that, as explained before, the 
turn-around scale factor has been evaluated as the time when the normalised radius reaches unity. For these two models 
(but not for generic dark energy models), one could determine it as the half time of the collapse process. 
While this is correct, we saw that it leads to numerical noise in $\zeta_{\rm ta}$, therefore we prefer not to consider 
the symmetries of the models. The turn-around is identical for both models for $z\gtrsim 2$, while for $\zeta_{\rm ta}$ 
this is the case at higher redshifts ($z\gtrsim 5$). 
Requiring higher initial overdensity to overcome a faster expansion, for a $\Lambda$CDM model $\zeta_{\rm ta}$ is 
higher than for an EdS model and this reflects in an earlier turn-around. 
We checked that the numerical solution for an EdS model agrees with the theoretical prediction.

\begin{figure}
 \begin{center}
  \includegraphics[scale=0.46,angle=-90]{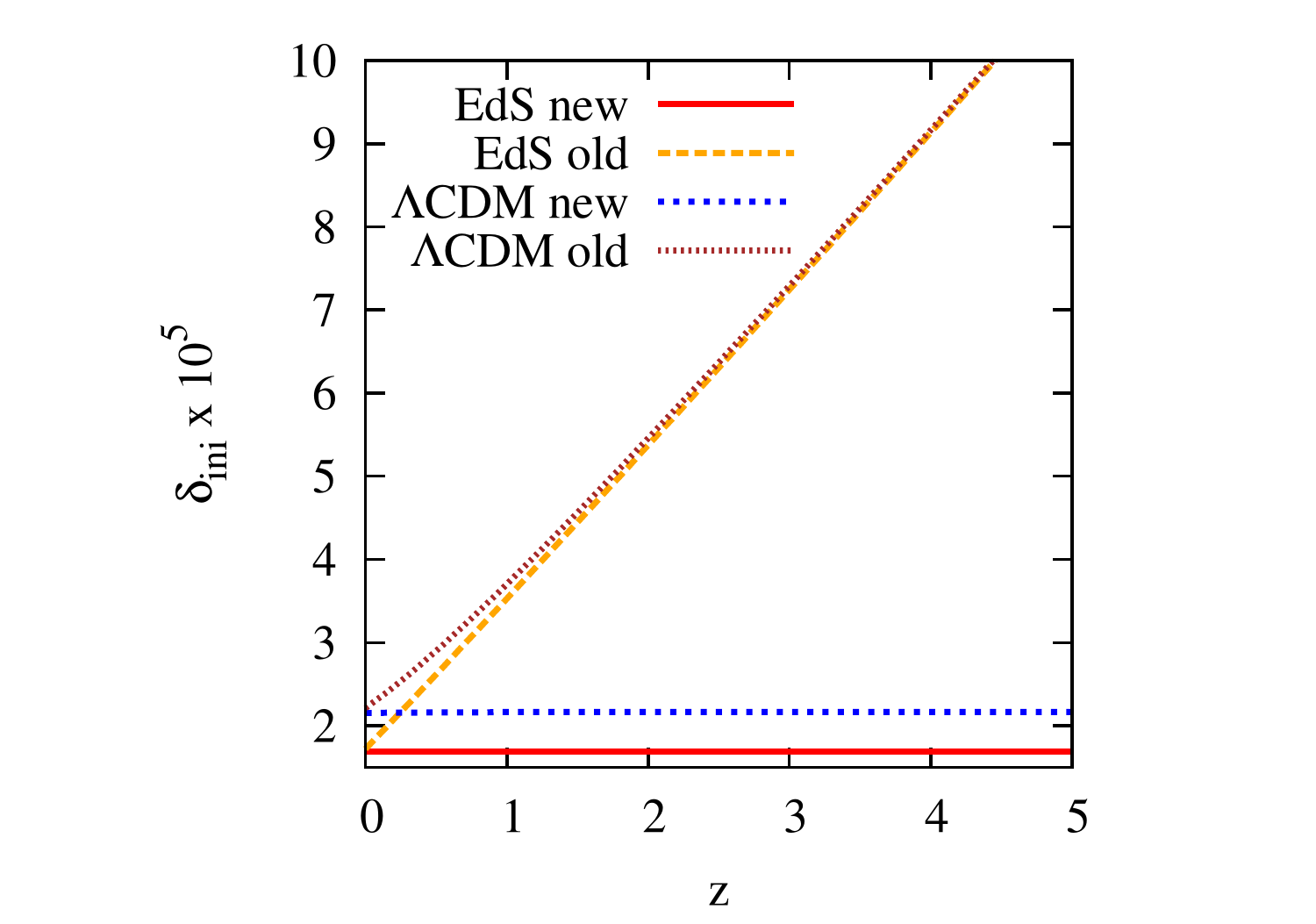}
  \includegraphics[scale=0.46,angle=-90]{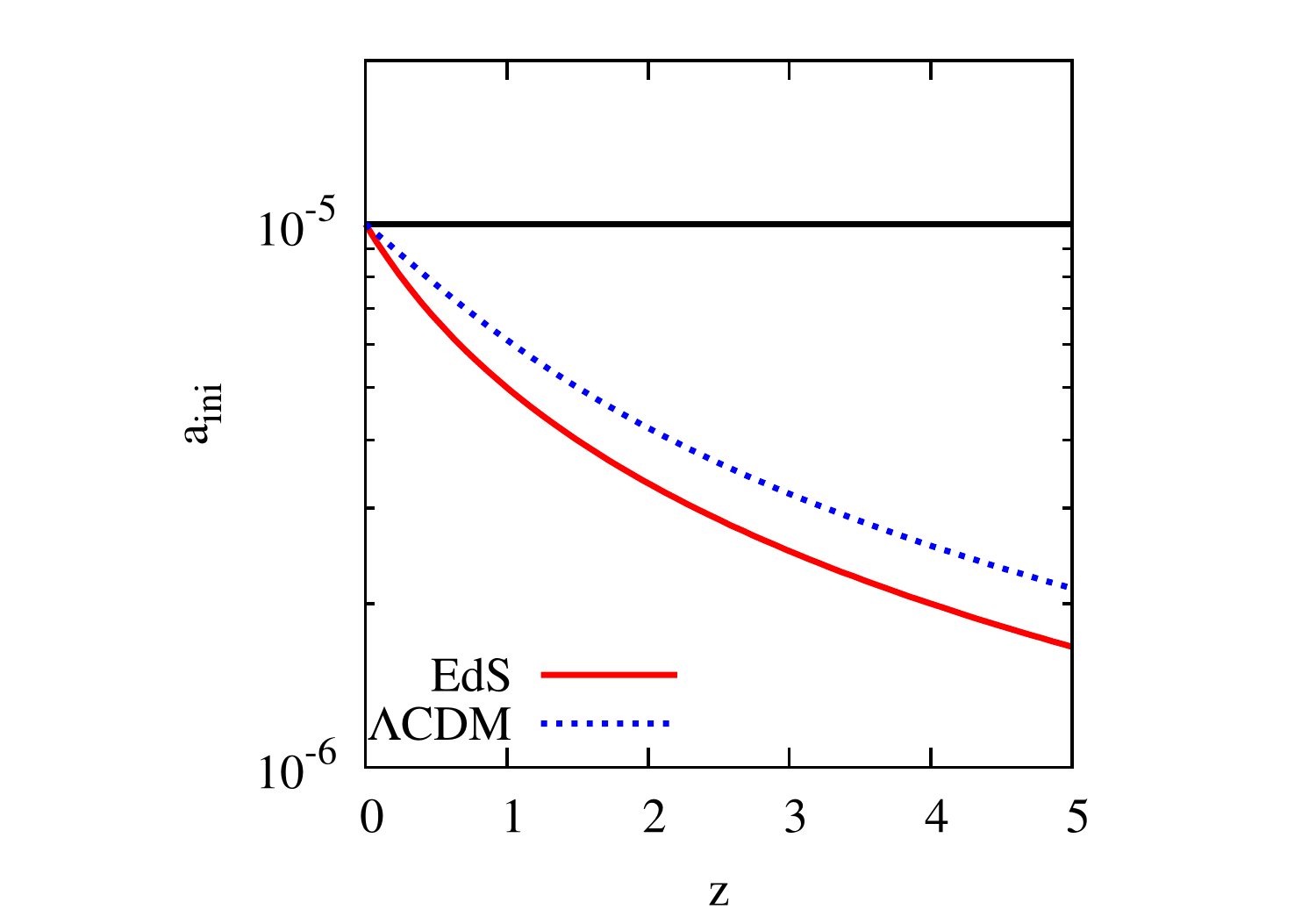}\\
  \includegraphics[scale=0.46,angle=-90]{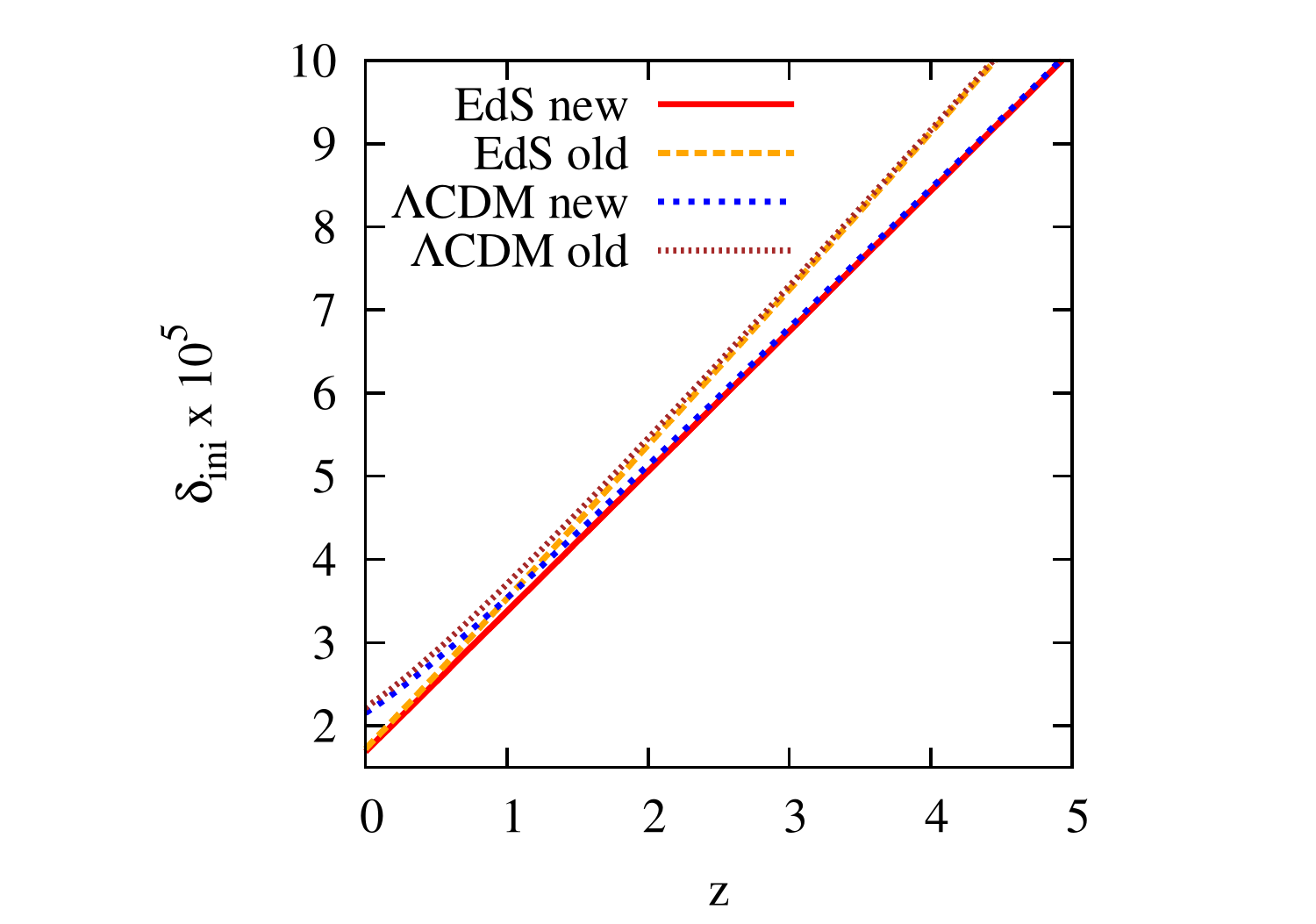}
  \caption{Top left panel: Evolution of the initial density perturbation $\delta_{\rm ini}$. The red solid (orange 
  dashed) curve represents the values for an EdS model for the new (old) approach; the blue short-dashed and the brown 
  dotted lines show the initial overdensity for a $\Lambda$CDM model for the new and the old implementation, 
  respectively. 
  Top right panel: Evolution of the initial scale factor $a_{\rm ini}$ where the integration starts. The solid red 
  curve is for the EdS model while the blue dashed one stands for the $\Lambda$CDM model. The solid line at $a=10^{-5}$ 
  reflects the old implementation. 
  Bottom panel: evolution of $\delta_{\rm ini}$ in the old and in the new implementation, evolving it from 
  $a_{\rm ini}$ to $a=10^{-5}$. Line styles and colours are as in the top panel.}
  \label{fig:ini}
 \end{center}
\end{figure}

Since the main difference between the old and the new implementation relies on a better estimation of the initial 
conditions, in Fig.~\ref{fig:ini} we show a comparison between the old and the new value of $\delta_{\rm ini}$ and how 
the initial scale factor $a_{\rm ini}$ where the integration starts is changing with respect to the collapse redshift. 
In the top left panel we show the initial overdensity effectively used by the code to integrate the equations of 
motion, while in the top right panel we show where the integration starts as a function of the collapse redshift. With 
the old initial conditions, we see, as expected, that the initial overdensity for the $\Lambda$CDM model is higher, due 
to the presence of the cosmological constant and for a collapse at $z=2$, they have essentially the same value of the 
EdS model. 
An earlier collapse implies of course a higher initial overdensity. In the new formulation, the $\delta_{\rm ini}$ for 
the $\Lambda$CDM model is still higher than for the EdS model, but now it is constant, regardless of the effective 
collapse time. This now translates into a change of $a_{\rm ini}$. While in the old formulation this was fixed to be 
$10^{-5}$, now we see that an earlier collapse corresponds to a smaller $a_{\rm ini}$. This is largely expected: the 
same overdensity, but at an earlier time will imply a faster and earlier collapse. This is the main difference which 
leads to a superior performance of the new implementation: instead of changing $\delta_{\rm ini}$, we change 
$a_{\rm ini}$ in a suitable way. Note also that to a lower $\delta_{\rm ini}$ corresponds a lower $a_{\rm ini}$ 
(compare the middle panel with the upper one).

It is interesting to have a better comparison between the implementations. To do so, we evolve the initial 
overdensity found with the new approach from $a_{\rm ini}$ to $a=10^{-5}$. Our results are presented in the bottom 
panel of Fig.~\ref{fig:ini}. While at $z=0$ the initial overdensity is approximately the same in both implementations, 
differences appear at higher collapse redshifts. Also in the new implementation the initial overdensity at $a=10^{-5}$ 
grows when the collapse redshift increases, but it does so at a smaller pace: this results in a bounded evolution for 
$\delta_{\rm c}$, as we saw for the $\Lambda$CDM model in Fig.~\ref{fig:deltacNew}. This result once again shows how 
crucial is the correct determination of the initial conditions.

So far we have assumed a specific driver from the \verb|GSL| library. We now investigate this aspect further taking 
into account the eventual stiffness of the equations and show the general good stability of the implementation.  
Extensive discussions about the theory and the numerical implementations, together with stability properties and 
accuracy for the explicit and the implicit class of Runge-Kutta methods can be found in 
\cite{Hairer1993,Hairer1996,Ascher1998} and formulas in \cite{Abramowitz1988}. The implicit Bulirsch-Stoer algorithm 
\verb|bsimp| is described in \cite{Bader1983} while for the Adams (\verb|msadams|) and BDF (\verb|msbdf|) multistep 
methods we refer the reader to \cite{Byrne1975,Brown1989,Hindmarsh2005}. 
For more details on the exact order and implementation of each single method, we refer the reader to the \verb|GSL| 
manual; here we merely refer to the meaning of each single solver. The \verb|bsimp| and \verb|msbdf| methods are 
suitable for stiff problems and are based on the Implicit Bulirsch-Stoer method and on a variable-coefficient linear 
multistep backward differentiation formula (BDF) method in Nordsieck form, respectively. Names of the solvers ending 
in {\it imp} represent implicit Runge-Kutta methods (\verb|rk1imp|, \verb|rk2imp|, \verb|rk4imp| of order 1, 2 and 4 
respectively), while the others are explicit methods (\verb|rk2| is an embedded Runge-Kutta (2, 3), \verb|rk4| a 
classical fourth order Runge-Kutta (RK), \verb|rkf45| an embedded Runge-Kutta-Fehlberg (4, 5), \verb|rkck| an embedded 
Runge-Kutta Cash-Karp (4, 5), \verb|rk8pd| an embedded Runge-Kutta Prince-Dormand (8, 9) method). 
The \verb|msadams| is a variable-coefficient linear multistep Adams method in Nordsieck form.

We run our code for several models: EdS, $\Lambda$CDM, early and oscillating dark energy models. Except for different 
running times, all the different solvers returned, within numerical precision, the same results, with differences 
smaller than one part in a million. This clearly shows that our implementation is stable and reliable and it is not 
influenced by the particular choice of the equation solver. Thanks to this, we can safely assume as default solver the 
embedded Runge-Kutta Cash-Karp (4, 5) \verb|rkck|.

In addition, our code is also extremely fast, especially for an EdS and a $\Lambda$CDM model, where, for coherence with 
the previous analysis on dark energy models, even if constant, we tabulated the equation of state of the cosmological 
constant. For models where the equation of state is either not a constant or a non-trivial function of the scale 
factor, due to the adaptive step size of the equation solver, the code becomes more slowly.\\
Using four threads on a Intel Core i5-2520M CPU at 2.50 GHz, the execution time was of 26.1 seconds to evaluate 512 
times $\delta_{\rm c}$ and $\Delta_{\rm V}$ for an EdS and a $\Lambda$CDM model at both the collapse and the 
virialization time at linearly equispaced steps between $z=0$ and $z=20$. Note that in the elapsed time, the following 
quantities were evaluated, but not asked in output: initial scale factor $a_{\rm ini}$ where the integration starts, 
evaluation of the initial overdensity $\delta_{\rm ini}$ via a root-finding and a Newton-Raphson algorithm (hence 
solving several times the full non-linear equations), creation of an internal table with $\delta$ as a function of $a$, 
determination of the turn-around scale factor $a_{\rm ta}$, of the overdensity at turn-around $\zeta_{\rm ta}$, the 
ratio between the virial and the turn-around radius $y_{\rm vir}=R_{\rm vir}/R_{\rm ta}$ and virial scale factor 
$a_{\rm vir}$. 
When solving the equations taking into account the shear and rotation invariants, the execution time is about 36.7 
seconds, with the same setup. The longer execution time is due to the additional computational burden for the 
evaluation of the growth factor.

Beside the EdS and the $\Lambda$CDM model, we tested our code against several dark energy models, in particular early 
dark energy and oscillating dark energy models. We found that, albeit the code needs a longer running time, the results 
are in perfect agreement with previous works and perfectly match the results of N-body simulations 
\citep[see][for a comparison between theory and simulations for early dark energy models]{Pace2010}. The main novelty 
is that now the linear overdensity parameter $\delta_{\rm c}$ is numerically more stable and bounded from above. We 
checked that up to $z=20$, $\delta_{\rm c}$ is stable and does not grow over the EdS value.

With respect to the setup for a $\Lambda$CDM model, these classes of models require greater care to suppress the 
numerical noise in $\delta_{\rm c}$. Two solutions are possible: the first one takes into account that for early dark 
energy models the evolution of the dark energy component is known analytically while for many oscillating dark energy 
models their equation of state leads to an exact analytic expression for $\Omega_{\rm de}(a)$. Therefore one could 
replace the tabulated expressions for $w(a)$ with the corresponding analytic expressions. The other solution, which we 
followed since in general we cannot expect to have an exact expression for $\Omega_{\rm de}(a)$, is to provide a very 
fine sample for the tabulated equation of state, increase the number of points used to evaluate the evolution of 
$\Omega_{\rm de}(a)$ and increase the accuracy of the code which evolves the perturbed equations of motion. 
Also note that the code will provide reliable results if the model under consideration will present a matter dominated 
era at early times, when the integration of the equations of motion starts.

\section{Conclusions}\label{sect:conclusions}
After reviewing the theory of the spherical collapse model, we discussed in detail the influence of each single 
parameter entering into the evaluation of the differential equations. With the standard set of equations for $\delta$ 
and $\theta$, we showed that the linear overdensity $\delta_{\rm c}$ is affected by numerical noise and it grows 
unbounded, while when using the equations for $f=1/\delta$ and $\theta$, the numerical noise is strongly suppressed, 
although the solution still grows without bounds. This behaviour is due to a combination of factors, namely the 
collapse only approximately reaches the numerical infinity and the initial conditions need to be adapted because of 
this.

Our main findings are:
\begin{itemize}
 \item The critical collapse density strongly depends on the value of the numerical infinity $\delta_{\infty}$, as 
       already pointed out by \cite{Herrera2017}. The numerical solution severely underestimates the analytic one for 
       small values of $\delta_{\infty}$ and it increases when this parameter is increased. We find convergence of 
       results for $\delta_{\infty}\geqslant 10^8$.
 \item After fixing $\delta_{\infty}$ to values high enough, we investigated how the result depends on the choice of 
       $a_{\rm ini}$, the initial time to start the integration of the equations of motion. Values too large 
       ($a_{\rm ini}\approx 10^{-3}$) lead the critical collapse density to growing quickly in the past, without any 
       clear asymptotic flattening. For a converged solution, we find that $a_{\rm ini}\leqslant 10^{-5}$.
 \item To suppress the numerical noise due to the numerical infinity, it is useful to work with $f=1/\delta$. When the 
       system collapses, $\delta\rightarrow\infty$ and $f\rightarrow 0$. In analogy to $\delta_{\infty}$, we 
       asked ourselves how small $\epsilon$ should be to reach numerical convergence. As expected, the smaller 
       $\epsilon$, the better is the numerical solution. For $\epsilon\leqslant 10^{-8}$, the solution has perfectly 
       converged.
\end{itemize}

These considerations led us to improve the determination of the initial conditions following these steps:
\begin{enumerate}
 \item Estimation of an initial slope for initial perturbations at $a_{\rm ini}=10^{-5}$.
 \item Given an arbitrary initial overdensity $\delta_{\rm ini}=1$, $a_{\rm ini}$ is scaled by the quantity 
       $\delta(a_{\rm c})/\delta(1)$.
 \item New estimation of the initial slope at the new $a_{\rm ini}$ and determination of $\delta_{\rm ini}$ leading to 
       the collapse at $a=a_{\rm c}$.
 \item Refinement of $\delta_{\rm ini}$ via a Newton-Raphson method.
 \item Once $\delta_{\rm ini}$ has been found, linear (non-linear) differential equations are started with appropriate 
       initial conditions: a linear (non-linear) relation is used to relate $\tilde{\theta}$ and $\delta$.
\end{enumerate}

In Figs.~\ref{fig:deltacNew} - \ref{fig:ata_zeta} we demonstrate the reliability and the power of the new 
implementation: the critical collapse density is now smooth and not affected by numerical noise, but most importantly, 
it is now bounded. We recover the theoretical results for the EdS model and show that we can deal with both the 
collapse and the virialization. The $\Lambda$CDM model correctly recovers the EdS solution at high redshifts. The 
code is stable and extremely fast as discussed before, and the results are independent of the particular solver 
adopted.

An interesting consequence of the new implementation is that instead of keeping fixed the starting time, this is 
shifted into the past and at the same time the initial overdensity is kept constant. This is clearly presented in the 
top and middle panels of Fig.~\ref{fig:ini}. The consequence of this is that, when compared to previous 
implementations, the initial overdensity at $a_{\rm ini}$ still grows as the redshift increases, but it does so at a 
smaller pace than before. In other words, the new initial overdensity is smaller than the initial overdensity in 
implementations A, B or C. A welcome effect is a bounded critical collapse density $\delta_{\rm c}$.

Since $\delta_{\rm c}$ (or equivalently $\nu=\delta_{\rm c}/D_{+}$) is the main ingredient of the mass function, it is 
important that it is evaluated precisely. As \cite{Herrera2017} discussed, the different implementations might not be 
of a concern now, but will be in the future with better data. We believe that our new approach is useful for studies in 
particular at high redshifts, as shown in \cite{Watson2013}.

Percent-level differences in the mass function will translate also to the determination of the matter and 
Sunyaev-Zel'dovich (SZ) power spectrum when the halo model is used \citep{Seljak2000,Ma2000a,Cooray2002}. This 
semi-analytic model is based on the assumption that all the matter is locked inside gravitationally bound spheres which 
can be treated as hard spheres. The total matter power spectrum is given by the sum of two terms: the 1-halo term is 
due to particle pairs belonging to the same halo and the 2-halo term depends on particle pairs belonging to two 
separate halos and therefore depends on the clustering properties of the halos. The 2-halo term dominates on large 
scales, the 1-halo term on small scales. Both terms require the evaluation of integrals over the mass function. 
According to the previous discussion, an unbounded $\delta_{\rm c}$ will not have any effect at small redshifts, but 
will underestimate by several percent the power spectrum at high redshifts, particularly for SZ studies. In addition, 
one has to take into account that a proper choice of the mass function at high redshift is required.

As we made clear, the goal behind this work is to solve the issues appearing when dealing with the collapse. One could 
pursue a simpler route by evaluating the initial conditions working with the turn-around, since densities are finite. 
To do so, one needs to know the turn-around time and in general saying that $t(a_{\rm c})=2t_{\rm ta}$ is a very good 
approximation, but it is not necessarily correct for generic models, although differences might be small. We preferred 
not to introduce any approximation into our code to arrive at correct and physically motivated results.

Finally, we would like to add few considerations to extensions of the code for more general models. For non-minimally 
coupled models such those described in \cite{Pace2014} and \cite{NazariPooya2016}, the code needs minimal modifications 
since the structure of the equations is the same as in Eqs.~\ref{eqn:fnl} and \ref{eqn:thetanl}. The only difference is 
in the term $\tfrac{3}{2}\Omega_{\rm m}(a)/f$ which needs to be multiplied by the appropriate function 
$G_{\rm eff}/G$, where $G_{\rm eff}$ is the effective gravitational constant of the model. If $G_{\rm eff}/G=4/3$, the 
code can be immediately used also for $f(R)$ models, as presented in \cite{Herrera2017}.

For clustering dark energy models, it is enough to add the two additional differential equations for $\delta_{\rm de}$ 
(Eq.~\ref{eqn:deltadeNL}) and $\tilde{\theta}_{\rm de}$ (Eq.~\ref{eqn:thetadeNL}). The main difference is that it is 
now more difficult to define the sphere radius. It is therefore easier to take into account that turn-around is defined 
as the time of maximum expansion, where the expansion of the perturbation halts. Safely assuming that dark energy 
perturbations are sub-dominant with respect to matter perturbations, one can solve the equation for $\tilde{\theta}$ 
and search for the root of the function $\tilde{\theta}+3$. To see this we follow the discussion in 
\cite{Abramo2009b}.\\
The physical distance and velocity are given by
\begin{equation*}
 \boldsymbol{r} = \left[a(t)+g(t,\boldsymbol{x}_0)\right]\boldsymbol{x}_0\;, \quad 
 \boldsymbol{v} = a(t)\left[H(t)\boldsymbol{x}_0+\boldsymbol{u}\right]\;,
\end{equation*}
where $g(t,\boldsymbol{x}_0)$ accounts for deviations from the background evolution. The perturbed velocity is given by 
the time derivative of $\boldsymbol{r}$ and defining $\dot{g}=a\dot{\tilde{g}}$, where $\dot{\tilde{g}}$ is the 
comoving peculiar velocity describing deviations from the Hubble flow, we can write
\begin{equation}
 \boldsymbol{v} = \left[H(t)+\dot{\tilde{g}}(t,\boldsymbol{x}_0)\right]\boldsymbol{r}_0\;, \quad 
 \boldsymbol{v} = h(t)\boldsymbol{r}_0\;.
\end{equation}
By comparing the previous expressions we can write
\begin{equation}
 h = H + \dot{\tilde{g}}\;, \quad \boldsymbol{u} = \dot{\tilde{g}}\boldsymbol{x}_0\;.
\end{equation}
where $h$ is the effective rate of expansion of the perturbed region. Taking the divergence of $\boldsymbol{u}$ we find
\begin{equation}
 \vec{\nabla}\cdot\boldsymbol{u} = \theta = 3\dot{\tilde{g}}+\boldsymbol{x}_0\cdot\vec{\nabla}\dot{\tilde{g}}\;,
\end{equation}
where, under the top-hat approximation, the last term vanishes. We can finally conclude that
\begin{equation}
 \frac{h}{H} = 1 + \frac{\tilde{\theta}}{3}\;,
\end{equation}
where $\tilde{\theta}=\theta/H$ as before. Turn-around implies $h=0$ and enforces a condition on $\tilde{\theta}$.

In addition, with the knowledge of $\delta_{\rm de}$, one can evaluate the dark energy equation of state inside the 
perturbed region \citep{Abramo2007,Abramo2008,Abramo2009b}:
\begin{equation}
 w_{\rm de}^{\rm c} = \frac{P_{\rm de}+\delta P_{\rm de}}{\rho_{\rm de}+\delta\rho_{\rm de}} 
                    = w_{\rm de} + \left(c_{\rm eff}^2-w_{\rm de}\right)\frac{\delta_{\rm de}}{1+\delta_{\rm de}}\;.
\end{equation}
This perturbed equation of state has two notable limits: when $\delta_{\rm de}\rightarrow0$ we recover the background 
solution $w_{\rm de}^{\rm c}=w_{\rm de}$, when $\delta_{\rm de}\gg 1$ we recover the effective sound speed, 
$w_{\rm de}^{\rm c}=c_{\rm eff}^2$. Note also that when $c_{\rm eff}=w_{\rm de}$, $w_{\rm de}^{\rm c}$ is always equal 
to the background value.

For clustering dark energy models one can finally evaluate the contribution of the dark energy mass to the total mass 
of the virialised structure $\epsilon_{\rm de}=M_{\rm de}/M_{\rm m}$ where $M_{\rm de}$ is the mass due to the dark 
energy component and $M_{\rm m}$ is the matter (dark matter plus baryon) mass. The matter mass is defined as
\begin{equation}
 M_{\rm m} = 4\pi\int_0^{R_{\rm vir}}dRR^2\rho_{\rm m}(1+\delta_{\rm m})\;,
\end{equation}
where $R_{\rm vir}$ is the virial radius. Under the top-hat approximation, this integral can be easily performed.

The dark energy mass comes in two flavours, according to how it is defined. In \cite{Creminelli2010}, $M_{\rm de}$ is 
defined as the mass associated with the dark energy perturbations only:
\begin{equation}
 M_{\rm de,p} = 4\pi\int_0^{R_{\rm vir}}dRR^2\rho_{\rm de}\delta_{\rm de}\;,
\end{equation}
but as explained in \cite{Batista2013} this definition will put matter and dark energy on unequal feet, therefore the 
authors modified the previous definition as in the following:
\begin{equation}
 M_{\rm de,T} = 4\pi\int_0^{R_{\rm vir}}dRR^2\rho_{\rm de}[1+3w_{\rm de}+(1+3c_{\rm eff})^2\delta_{\rm de}]\;,
\end{equation}
which is a direct consequence of the generalised Poisson equation $\nabla^2\psi=4\pi G(\rho+3P/c^2)$. A more physical 
reason is that gravitational lensing depends on the total mass of the perturbation. The value of $\epsilon$ is at 
most at the percent level (for models with $c_{\rm eff}^2=0$) and it is important only at late times, when dark energy 
perturbations become more important. 
To evaluate $\epsilon_{\rm de}$ we integrate the non-linear equations for $\delta_{\rm de}$ and $\delta_{\rm m}$ till 
$a_{\rm v}$.

A further point to consider for clustering dark energy models is that the evolution of $\delta_{\rm de}$ strongly 
depends on its equation of state. At early times, $\delta_{\rm de}\propto(1+w_{\rm de})\delta_{\rm m}$, therefore for 
phantom models, $\delta_{\rm de}<0$. In other words, to matter overdensities correspond dark energy underdensities 
(voids) and vice versa; therefore a cut-off must be set to the evolution of $\delta_{\rm de}$, such that 
$\delta_{\rm de}\geqslant -1$. 
Similar considerations can be applied to interacting dark energy models.

Finally, when adding shear and rotation, it is only necessary to have a prescription for their evolution. 
For the model by \cite{Reischke2016b}, this requires at each time step to calculate the linear growth factor. 
Since for a generic dark energy model there is no analytic solution, the computation of the equations of the spherical 
collapse becomes computationally demanding. This can be easily avoided by tabulating at the beginning the values of the 
linear growth factor of the model and interpolate them at the required time.

\section*{Acknowledgements}
FP acknowledges support from the post-doctoral STFC grant R120562 'Astrophysics and Cosmology Research within the JBCA 
2017-2020'. SM acknowledges support from the German DFG grant BA 1369/20-2. 
The authors thank Robert Reischke, Bj\"orn Sch\"afer and Lucas Lombriser for useful discussions.

\bibliographystyle{JHEP}
\bibliography{Implementation.bbl}

\appendix
\section{Code structure}\label{sect:code}
For completeness, here we outline the structure of our novel implementation in more detail, presenting it as a 
pseudo-code.

\begin{algorithm*}[ht]
 \DontPrintSemicolon
 \SetAlgoLined
 \SetStartEndCondition{ (}{)}{)}\SetAlgoBlockMarkers{}{\}}%
 \AlgoDisplayBlockMarkers%
 
 \SetKwProg{Fn}{}{\{}{}\SetKwFunction{cosmo}{void cosmology}%
 \SetKwProg{Fn}{}{\{}{}\SetKwFunction{scm}{void spcModel}%
 
 \TitleOfAlgo{Code for the spherical collapse model}
 
 \vspace{0.4cm}
 
 \Fn(\tcc*[h]{Initialise the cosmological model}){\cosmo{$\Omega_{\rm m,0}$,$\Omega_{\rm de,0}$,$h$,$w_{\rm de}(a)$}}{
 
 \vspace{0.1cm}
 
 \KwData{Initialization of internal variables}
 
 \vspace{0.1cm}
 
 \tcc{Internal calculations}
 
 \vspace{0.1cm}
 
 $g(a) \gets \int_1^a (1+w_{\rm de}(x))~d\ln{x}$ \tcc*[r]{Dark energy evolution}
 
 \vspace{0.1cm}
 
 $H(a) \gets H_0\sqrt{\Omega_{\rm m,0}a^{-3}+(1-\Omega_{\rm m,0}-\Omega_{\rm de,0})a^{-2}+\Omega_{\rm de,0}g(a)}$ 
 \tcc*[r]{Hubble function}
 
 \vspace{0.1cm}
 
 $H^{\prime}(a) \gets -H_0^2[3\Omega_{\rm m,0}a^{-3}+2(1-\Omega_{\rm m,0}-\Omega_{\rm de,0})a^{-2}+
 3\Omega_{\rm de,0}(1+w_{\rm de}(a))g(a)]/2/a/H(a)$ 
 \tcc*[r]{Derivative of the Hubble function}
 }
 
 \vspace{0.4cm}
 
 \Fn(\tcc*[h]{Initialise the spherical collapse model}){\scm{$z_{\rm c}$, $\sigma^2-\omega^2$, flags}}{
 
 \vspace{0.1cm}
 
 \KwData{Initialization of internal variables}
 
 \vspace{0.1cm}
 
 $a_{\rm ini} \gets 10^{-5}$ \tcc*[r]{Maximum initial scale factor}
 
 \vspace{0.1cm}
 
 \tcc{Get the right initial conditions: $a_{\rm ini}$, $\delta_{\rm ini}$}
 
 \vspace{0.1cm}
 
 $x^2+\left[2+a_{\rm ini}H^{\prime}(a_{\rm ini})/H(a_{\rm ini})\right]x-3\Omega_{\rm m}(a_{\rm ini})/2=0$
 \tcc*[r]{Initial slope $\delta_{\rm ini}\propto a^n$}
 
 \vspace{0.1cm}
 
 $\delta_{\rm ini} \gets 1$ \tcc*[r]{Arbitrary initial condition for $\delta_{\rm lin}$}
 
 \vspace{0.1cm}
 
 $a_{\rm ini} \gets a_{\rm ini}\delta_{\rm lin}(z_{\rm c})/\delta_{\rm lin}(z_{\rm c}=0)$ 
 \tcc*[r]{Initial scale factor}
 
 \vspace{0.1cm}
 
 $x^2+\left[2+a_{\rm ini}H^{\prime}(a_{\rm ini})/H(a_{\rm ini})\right]x-3\Omega_{\rm m}(a_{\rm ini})/2=0$
 \tcc*[r]{Initial slope for $\delta_{\rm ini}\propto a^n$}
 
 \vspace{0.1cm}
 
 $\delta_{\rm ini} \gets 1/\delta_{\rm NL}(\delta_{\rm ini})\rightarrow 0$
 \tcc*[r]{Initial overdensity $\sim 10^{-5}$ via root finding}
 
 \vspace{0.1cm}
 
 \tcc{Get useful quantities $a_{\rm ta}$, $\zeta_{\rm ta}$, $y_{\rm vir}$, $a_{\rm vir}$}
 
 \vspace{0.1cm}
 
 $a_{\rm ta}=$ findAt() \tcc*[r]{Turn-around scale factor}
 
 \vspace{0.1cm}
 
 $\zeta_{\rm ta}=\delta_{\rm NL}(a_{\rm ta})+1$ \tcc*[r]{Find overdensity at turn-around}
 
 \vspace{0.1cm}
 
 $y_{\rm vir}=R_{\rm vir}/R_{\rm ta}$ \tcc*[r]{Use \cite{Wang1998,Maor2005,Maor2007}}
 
 \vspace{0.1cm}
 
 $a_{\rm vir}=$ findAv() \tcc*[r]{Virial scale factor}
 }
\end{algorithm*}

\begin{algorithm*}[ht]
 \DontPrintSemicolon
 \SetAlgoLined
 \SetStartEndCondition{ (}{)}{)}\SetAlgoBlockMarkers{}{\}}%
 \AlgoDisplayBlockMarkers%
 
 \TitleOfAlgo{Necessary routines for the spherical collapse model}

 \SetKwProg{Fn}{}{\{}{}\SetKwFunction{at}{double findAt}%
 \SetKwProg{Fn}{}{\{}{}\SetKwFunction{yv}{double findYv}%
 \SetKwProg{Fn}{}{\{}{}\SetKwFunction{av}{double findAv}%
 \SetKwProg{Fn}{}{\{}{}\SetKwFunction{dt}{double delta\_t}%
 \SetKwProg{Fn}{}{\{}{}\SetKwFunction{dc}{double delta\_c}%
 \SetKwProg{Fn}{}{\{}{}\SetKwFunction{DV}{double Delta\_V}%

 \Fn(\tcc*[h]{Determine the turn-around scale factor $a_{\rm ta}$}){\at{}}{
 
 \vspace{0.1cm}

 $a_{\rm ta} \gets \tilde{\theta}_{\rm NL}(a)=-3$ 
 \tcc*[r]{Use Eq.~\ref{eqn:thetanl} with a root-finding algorithm}
 }
 
 \vspace{0.35cm}
 
 \Fn(\tcc*[h]{Determine $y_{\rm vir}=R_{\rm vir}/R_{\rm ta}$}){\yv{}}{
 
 \vspace{0.1cm}
 
 \Case(\tcc*[h]{Choose between different virialization recipes}){virialization condition}{
 \tcc{Shown for \cite{Wang1998,Maor2005,Maor2007}}

 $\eta_{\rm t} \gets 2\zeta_{\rm ta}^{-1}\Omega_{\rm de}(a_{\rm ta})/\Omega_{\rm m}(a_{\rm ta})$
 
 \vspace{0.1cm}
 
 $\eta_{\rm v} \gets 2\zeta_{\rm ta}^{-1}\Omega_{\rm de}(a_{\rm c})/\Omega_{\rm m}(a_{\rm c})(a_{\rm at}/a_{\rm c})^3$
 
 \vspace{0.1cm}
 
 $y_{\rm vir} \gets (1-\eta_{\rm v}/2)/(2+\eta_{\rm t}-3\eta_{\rm v}/2)$
 }
 }
 
 \vspace{0.35cm}
 
 \Fn(\tcc*[h]{Determine the virialization scale factor $a_{\rm vir}$}){\av{}}{
 
 \vspace{0.1cm}
 
 $a^0 \gets a_{\rm c}$
 
 \Repeat{$\left|(a^{(n+1)}-a^{(n)})/a^{(n)}\right| \leq tol$}{
 \vspace{0.1cm}
 
 $y(a)= a^{(n)}\left\{\zeta_{\rm ta}/[1+\delta_{\rm NL}(a^{(n)})]\right\}^{1/3}/a_{\rm ta}$ 
 \tcc*[r]{To be used with a root-finding algorithm}
 
 \vspace{0.1cm}
 
 $a(y_\mathrm{vir}^{(n)}) \gets y(a)-y_\mathrm{vir}^{(n)}=0$
 
 \vspace{0.1cm}
 
 $a^{(n)} \overset{{\rm Eq}.~(\ref{eqn:vir_eq})}{\longrightarrow} y_{\mathrm{vir}}^{(n)}
 \longrightarrow a\left(y_\mathrm{vir}^{(n)}\right)=a^{(n+1)}$}
 }
 
 \vspace{0.35cm}
 
 \Fn(\tcc*[h]{Determine $\delta_{\rm lin}$ at turn-around}){\dt{}}{
 
 \vspace{0.1cm}
 
 $\delta \gets \delta_{\rm lin}(a_{\rm ta})$ \tcc*[r]{Use Eq.~\ref{eqn:fl}, $f \gets 1/\delta_{\rm lin}$}
 }
 
 \vspace{0.35cm}
 
 \Fn(\tcc*[h]{Determine $\delta_{\rm c}$ at collapse and virialization}){\dc{}}{
 
 \vspace{0.1cm}
 \tcc{Use Eq.~\ref{eqn:fl}, $f \gets 1/\delta_{\rm lin}$}
 
 $\delta_{\rm c}(a_{\rm c}) \gets \delta_{\rm lin}(a_{\rm c})$
 
 $\delta_{\rm c}(a_{\rm vir}) \gets \delta_{\rm lin}(a_{\rm vir})$
 }
 
 \vspace{0.35cm}
 
 \Fn(\tcc*[h]{Determine $\Delta_{\rm V}$ at collapse and virialization}){\DV{}}{
 
 \vspace{0.1cm}
 
 $\Delta_{\rm V}(a_{\rm c}) \gets \zeta_{\rm ta}\Omega_{\rm m}(a_{\rm c})(a_{\rm c}/a_{\rm ta})^3/y_{\rm vir}^3$
 
 $\Delta_{\rm V}(a_{\rm vir}) \gets \zeta_{\rm ta}\Omega_{\rm m}(a_{\rm vir})(a_{\rm vir}/a_{\rm ta})^3/y_{\rm vir}^3$
 }
\end{algorithm*}

\label{lastpage}

\end{document}